\documentclass{article}

\usepackage{authblk}
\usepackage{hyperref}
\usepackage{graphics}
\usepackage{amsmath}
\usepackage{mathtools}
\usepackage{xcolor} 
\usepackage{hyperref}
\usepackage[perpage]{footmisc}
\usepackage[margin=1in]{geometry}

\providecommand{\keywords}[1]
{
  \small	
  \textbf{\textit{Keywords---}} #1
}

\title{First operation of the FAMU experiment at the RIKEN-RAL high intensity muon beam facility}

\author[1]{A.~Adamczak}
\author[2]{D.~Bakalov}
\author[3,4]{G.~Baldazzi} 
\author[5]{M.~Baruzzo} 
\author[6,7]{R.~Benocci} 
\author[6]{R.~Bertoni} 
\author[6,8]{M.~Bonesini}
\author[9,10]{S.~Capra} 
\author[5]{D.~Cirrincione} 
\author[6,8]{M.~Clemenza} 
\author[11,12]{L.~Colace}  
\author[5,13]{M.~Danailov} 
\author[2]{P.~Danev}
\author[14,15]{A.~de~Bari}
\author[15]{C.~De~Vecchi} 
\author[4]{D.~Di~Ferdinando} 
\author[16,17]{E.~Fasci} 
\author[6]{R.~Gaigher} 
\author[16,17]{L.~Gianfrani} 
\author[18]{A.~D.~Hillier} 
\author[19]{K.~Ishida} 
\author[18]{J.~S.~Lord} 
\author[14,15]{A.~Menegolli\footnote{Corresponding author. e-mail: alessandro.menegolli@unipv.it}}
\author[5]{E.~Mocchiutti}
\author[5,20]{S.~Monzani} 
\author[16,17]{L.~Moretti} 
\author[3,21]{G.~Morgante} 
\author[5]{C.~Pizzolotto} 
\author[9,10]{A.~Pullia} 
\author[15,22]{M.~Pullia} 
\author[9,23]{R.~Ramponi} 
\author[6,8]{H.~E.~Roman} 
\author[15]{M.~Rossella} 
\author[14,15]{R.~Rossini} 
\author[3,4]{A. Sbrizzi} 
\author[2]{M.~Stoilov} 
\author[5]{J.~J.~Su\'arez-Vargas\footnote{Now at INFN LNS}}
\author[24]{G.~Toci} 
\author[11]{L.~Tortora} 
\author[6]{E.~Vallazza} 
\author[18]{K.~Yokoyama} 
\author[5,19,20]{A.~Vacchi}

\affil[1]{Institute of Nuclear Physics, Polish Academy of Sciences, Radzikowskiego 152, PL31342 Krak\'{o}w, Poland}
\affil[2]{Institute for Nuclear Research and Nuclear Energy,
Bulgarian Academy of Sciences, blvd.\ Tsarigradsko ch.~72, Sofia 1784, Bulgaria}
\affil[3]{Dipartimento di Fisica e Astronomia ``A. Righi", Universit\`a di Bologna, viale Berti Pichat 6/2, Bologna, Italy}
\affil[4]{Sezione INFN di Bologna, viale Berti Pichat 6/2, Bologna, Italy}
\affil[5]{Sezione INFN di Trieste, via A. Valerio 2, Trieste, Italy} 
\affil[6]{Sezione INFN di Milano Bicocca, Piazza della Scienza 3, Milano, Italy}
\affil[7]{Dipartimento di Scienze dell'Ambiente e della Terra, Universit\`a di Milano Bicocca, Piazza della Scienza 1, Milano, Italy}
\affil[8]{Dipartimento di Fisica ``G. Occhialini", Universit\`a di   Milano Bicocca, Piazza della Scienza 3, Milano, Italy}
\affil[9]{Sezione INFN di Milano, via Celoria 16, Milano, Italy}
\affil[10]{Dipartimento di Fisica, Universit\`a degli Studi di Milano, via Celoria 16, Milano, Italy}
\affil[11]{Sezione INFN di Roma Tre, Via della Vasca Navale 84, Roma, Italy}
\affil[12]{Dipartimento di Ingegneria, Universit\`a degli Studi Roma Tre, Via V. Volterra 62, Roma, Italy} 
\affil[13]{Sincrotrone Elettra Trieste, SS14, km 163.5, Basovizza, Italy}
\affil[14]{Dipartimento di Fisica ``A. Volta", Universit\`a di Pavia, via A.~Bassi 6, Pavia, Italy}
\affil[15]{Sezione INFN di Pavia, Via A.~Bassi 6, Pavia, Italy}
\affil[16]{Sezione INFN di Napoli, Via Vicinale Cupa Cintia 26, Napoli, Italy}
\affil[17]{Dipartimento di Matematica e Fisica ``L. Vanvitelli”, Universit\`a della Campania, Viale Lincoln 5, Caserta, Italy}
\affil[18]{ISIS Neutron and Muon Source, STFC Rutherford-Appleton Laboratory, Didcot, OX11 0QX, United Kingdom}
\affil[19]{Riken Nishina Center, RIKEN, 2-1 Hirosawa, Wako, Saitama 351-0198, Japan}
\affil[20]{Dipartimento di Scienze Matematiche, Informatiche e Fisiche, Universit\`a di Udine, via delle Scienze 206, Udine, Italy}
\affil[21]{INAF-OAS Bologna, via P.~Gobetti 93/3, Bologna, Italy}
\affil[22]{Centro Nazionale di Adroterapia Oncologica (CNAO), Via Borloni 1, Pavia, Italy} 
\affil[23]{IFN-CNR, Dipartimento di Fisica, Politecnico di Milano, piazza Leonardo da Vinci 32, Milano, Italy}
\affil[24]{INO-CNR, via Madonna del Piano 10, 50019 Sesto Fiorentino, Italy}

\begin{document}

\maketitle

\begin{abstract}
The FAMU experiment, supported and funded by the Italian Institute of Nuclear Physics (INFN) and by the Science and Technology Facilities Council (STFC), aims to perform the first measurement of the ground-state hyperfine splitting ({\em 1S-hfs}) of muonic hydrogen ($\mu H$). This quantity is highly sensitive to the proton's Zemach radius $R_Z$. An experimental determination of $R_Z$ provides significant constraints on the parametrization of the proton form factors as well as on theoretical models describing the proton’s electromagnetic structure. Following years of technological and methodological development, the FAMU experiment began operations in 2023 at Port 1 of the RIKEN-RAL muon beam line at the ISIS Neutron and Muon Source facility (Didcot, UK). In this paper, we first describe the unique detection technique employed by FAMU to determine the {\em 1S-hfs} of muonic hydrogen, followed by a detailed presentation of the final experimental layout. Finally, we report the first outcome from the 2023 commissioning run and from the initial physics runs performed in 2023 and 2024.    
\end{abstract}

\keywords{Nuclear reactions; muon-induced reactions; exotic atoms and molecules; muonic atoms and molecules; optics; laser optical systems: design and operation}

\newpage

\section{Introduction}
\label{sec:intro}

The results coming from the experiment on Lamb shift in muonic hydrogen $\mu H$ which allowed derivation of the proton charge radius with unprecedented precision~\cite{bib:antognini2013}, contradicted measurements done with different techniques~\cite{bib:pohl2013}, extracting the proton charge radius from electronic hydrogen or elastic electron-proton scattering. This experimental result gave rise to the iconic {\em proton radius puzzle} and sparked a decade-long flurry
of theoretical and experimental analyses, all of which agreed that additional research into the muonic hydrogen atom system was necessary. In fact, the energy levels of muonic hydrogen are orders of magnitude more sensitive to the details of the proton structure than the levels of normal hydrogen, since the reduced mass of the nucleus-muon system is 187 times bigger than the one of the nucleus-electron system. This long-chased laser spectroscopy measurement put in foreground the sensitivity of the muonic hydrogen system taken as an observatory able to serve, depending on the transition considered and the precision attained, as a cross check of predictions in
fields like QED, nuclear and particles physics. 
This occurrence immediately revived the previous idea of using laser spectroscopy
to measure the hyperfine splitting ({\em 1S-hfs}) in the ground state of muonic hydrogen~\cite{bib:bakalov1993}~\cite{bib:dupays2003}.
The hyperfine splitting in muonic hydrogen represents a case where the accuracy of
QED calculations exceeds the accuracy of the known values of fundamental physical
parameters. Hence the measurement of $\Delta E ^{1S}_{hfs}$ provides a unique possibility for the
measurement of the proton magnetic structure with higher accuracy than
that achievable in nuclear or particle physics experiments.
The Zemach radius $R_Z$ is the physical quantity related to the electromagnetic
properties of the proton that can be extracted from the muonic hydrogen {\em 1S-hfs} measurements. It can
be expressed as:

\begin{equation}
R_Z = \frac{\Delta E^{1S-hfs}_{exp}/E^F - 1 - \delta^{QED} - \delta^{recoil} - \delta^{pol} - \delta^{hvp}}{2m_{\mu p}\alpha}
\end{equation}
where $m_{\mu p}$ is the reduced mass of muonic hydrogen. $E_F$ is the Fermi energy expressed in terms of muon and proton masses $m_{\mu}$, $m_p$, and of the magnetic dipole moment of the proton $\mu_p$:

\begin{equation}
E^F = \frac{8}{3} \alpha^4 c^2 \frac{m^2_\mu m^2_p}{(m_\mu + m_p)^3} \mu_p
\end{equation}
while $\delta^{QED}, \delta^{recoil}, \delta^{pol}$ and $\delta^{hvp}$ are correction terms related to the proton electromagnetic structure and to the strong interaction~\cite{bib:antognini2022}. The Zemach radius of the proton $R_Z$ is defined in terms of the first moment of the convolution of the charge $\rho_E({\bf r})$ and magnetic moment distributions $\rho_M({\bf r})$~\cite{bib:zemach56}:

\begin{equation}
R_Z = \int d^3{\bf r} |{\bf r}| \int d^3{\bf r'} \rho_E ({\bf r'}) \rho_M({\bf r} - {\bf r'})
\end{equation}
The hyperfine splitting of $\mu H$ is the quantity that is most sensitive to the proton's Zemach radius $R_Z$. The experimental value of $R_Z$ places significant constraints on the parametrization of proton form factors as well as the theoretical models of proton electromagnetic structure~\cite{bib:antognini2022}~\cite{bib:pascalutsa2021}; the measurement of the {\em 1S-hfs} in the $\mu H$ system is therefore essential~\cite{bib:carlson2011}~\cite{bib:karshenboim2005}. A method  was originally drawn up at a time when the laser system could not be produced with the necessary performance~\cite{bib:bakalov1993}. A possible approach~\cite{bib:adamczak2001} was subsequently identified by the FAMU (Fisica Atomi MUonici - Physics of Muonic Atoms) collaboration, supported and funded by the Italian Institute of Nuclear Physics (INFN)~\cite{bib:adamczak2012}~\cite{bib:pizzolotto2020}.

The paper is organized as follows: in Sec.~\ref{sec:experiment} the technique proposed by the FAMU experiment to measure the {\em 1S-hfs} in the $\mu H$ system is presented, highlighting previous achievements of the FAMU Collaboration towards the definition of the final setup of the experiment, which is illustrated in all its components in Sec.~\ref{sec:setup}. The first operation of the FAMU experiment, which took place in 2023 at the Port 1 of the RIKEN-RAL muon beam at the ISIS Neutron and Muon Source facility (Didcot, UK), is presented in Sec.~\ref{sec:operations}, with a particular emphasis put into the commissioning phase of July 2023 (Sec.~\ref{subsec:commissioning}) and on the first runs for physics of 2023 and 2024 (Sec.~\ref{subsec:physics}). Conclusions and perspectives are finally summarized in Sec.~\ref{sec:conclusions}.

\section{The FAMU experiment}
\label{sec:experiment}

\begin{figure*}[htpb]
\centering
\includegraphics[width=.8\linewidth]{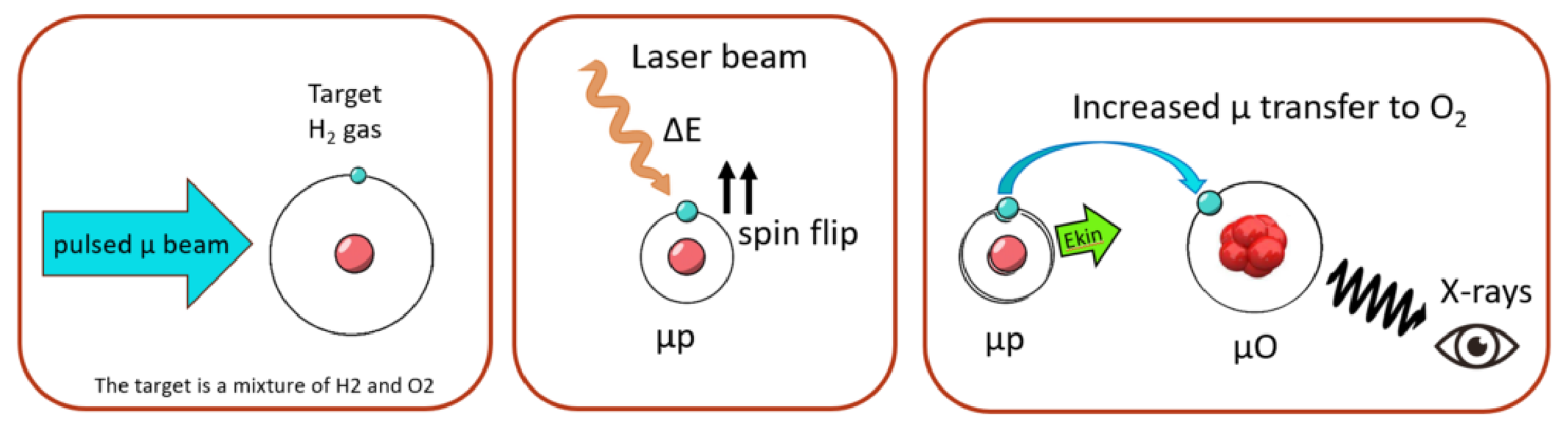}
\caption{Scheme of the FAMU method to excite the {\em 1S-hfs} transition in muonic hydrogen through a MIR laser beam, with the subsequent transfer of the muon to molecular Oxygen and the detection of the typical $\mu$O X-rays as a signature of the transition.}
\label{fig:method}       
\end{figure*}

\begin{figure*}[htpb]
\centering
\includegraphics[width=.8\linewidth]{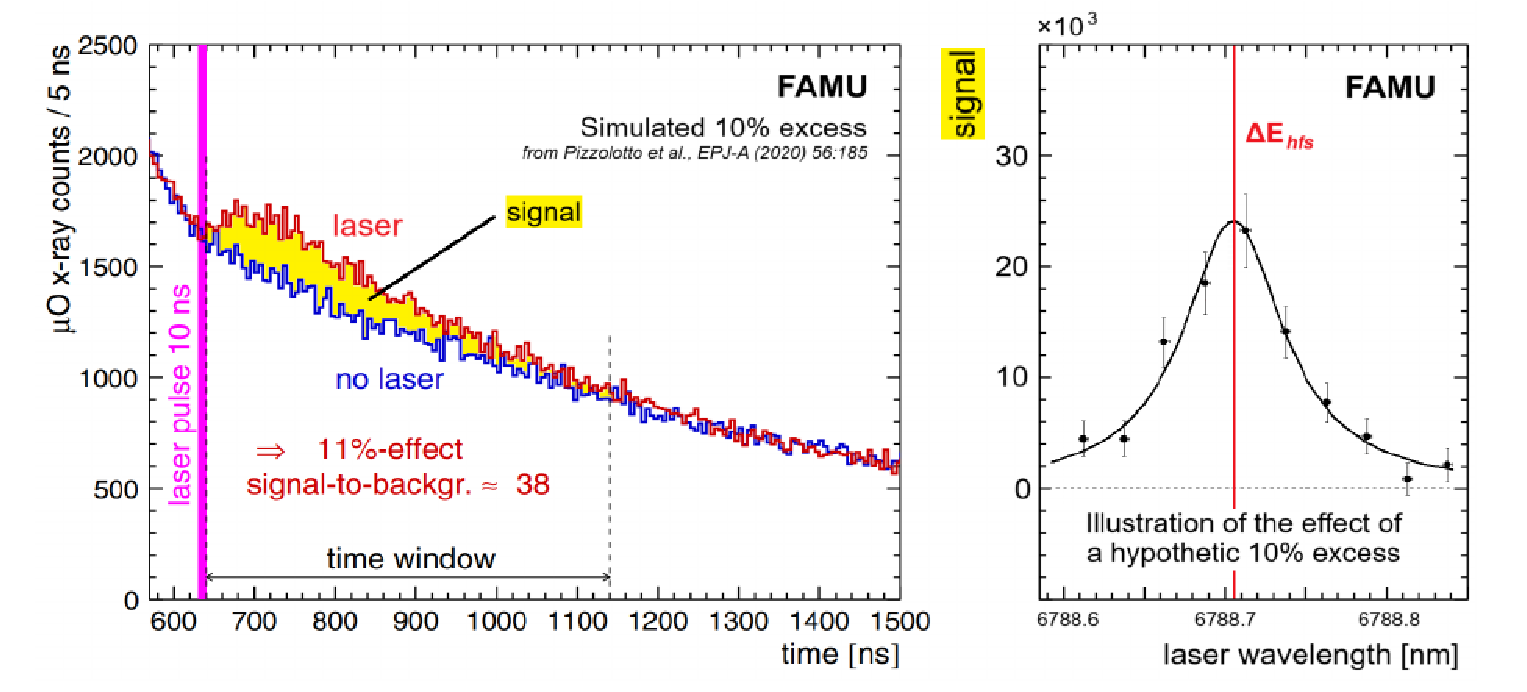}
\caption{Behaviour of the FAMU observable, the excess of delayed $\mu$O X-rays.
Left: toy simulation of a hypothetical $\sim$10\% effect on the FAMU observable (yellow coloured area). Right: illustration of a hypothetical resonance of the FAMU observable as a function of the laser wavelength, where the mean value is the resulting estimation of $\Delta E_{1S-hfs}$ (the value chosen here as $\Delta E_{1S-hfs}$ is the latest theoretical prediction, not a FAMU experimental result).}
\label{fig:resonance}       
\end{figure*}

Different experimental proposals for the measurement of $\Delta E_{1S-hfs}$ in the $1S$ state of the $\mu H$ system have been put forth in recent years to provide top-accuracy data on the Zemach radius of the proton~\cite{bib:kanda2018}~\cite{bib:amaro2022}~\cite{bib:adamczak2018}. This has been motivated by the need for new data on the proton electromagnetic structure that, as we have seen, had become a problem with the proton charge radius determination from the Lamb shift in muonic hydrogen. The laser excitation of the ortho $F = 1$ hyperfine sub-level of the ground state of the muonic hydrogen atom from the para $F = 0$ sub-level is the challenging task. This is a very weak $M1$ magnetic dipole transition with probability $P$ of only 

\begin{equation}
    P = 2 \cdot 10^{-5} (E[J]) (S[m^2])^{-1} (T[K])^{-1/2}
    \label{eq:probability}
\end{equation}
with $E$ laser pulse energy, $S$ laser beam cross section and $T$ target temperature~\cite{bib:adamczak2012}. In order to increase the likelihood of this laser-induced transition to a detectable level, an optical multi-pass cavity must be used. The muonic hydrogen atom is excited from the ground singlet to the triplet state with a laser, tunable around the resonance frequency $\Delta E_{1S-hfs}/h \sim$ 44 THz. 
The experimental method, adopted by the FAMU Collaboration, uses a peculiar characteristic of the muon transfer between muonic hydrogen and oxygen molecules, to detect when the muonic hydrogen are excited by the laser. For this reason, at first muonic hydrogen atoms are formed in a mixture of oxygen and hydrogen and then interact with a laser radiation tuned to a frequency around the hyperfine transition resonance, see Fig.~\ref{fig:method}. Finally, collisions of $\mu H$ with oxygen lead to the reaction:

\begin{equation}
 \mu H + O \rightarrow \mu O^* + H
\end{equation}
The observable quantity used as signature of this reaction is the time distribution of the muon transfer events from hydrogen to oxygen~\cite{bib:adamczak2018}, which are identified through the characteristic X-rays emitted during the de-excitation of the muonic oxygen. The maximum deviation of this X-ray time distribution from the time distribution in absence of the laser indicates that the laser source is tuned at the resonance frequency, see Fig.~\ref{fig:resonance}.
The $\mu H$ atoms that have been excited to the triplet state with the laser pulse are accelerated, after the de-excitation in subsequent collisions with the surrounding $H_2$ molecules, by nearly 0.12~eV; the atoms carry the released energy away as kinetic energy. Since the rate of muon transfer to oxygen $\lambda_{pO}(E)$ varies with the $\mu H$ kinetic energy $E$~\cite{bib:stoilov2023}, the observed time distribution of the characteristic X-rays is perturbed as compared to the time distribution in the absence of laser radiation; the resonance frequency is recognized by the maximal response of the X-ray time distribution. The efficiency of this method of detecting the events of laser-induced hyperfine excitation of the $\mu H$ depends on how much the rate of muon transfer from accelerated $\mu H$ atoms exceeds the transfer rate from thermalized atoms. The hydrogen-oxygen mixture had been selected for the FAMU method because of the evidence for a sharp energy dependence of $\lambda_{pO}(E)$ at thermal and near epithermal energies (nearly an order of magnitude)~\cite{bib:stoilov2023}, that is not observed in other gases.

High precision spectroscopy is required to accurately identify this extremely weak transition signal and, in order to recover the signal, all potential background sources must be reduced to a minimum. This called for thorough simulations before building the final layout. The pulsed muon beam will penetrate the cryogenic target full of high purity hydrogen with low oxygen contamination, primarily forming muonic hydrogen that quickly de-energizes and reaches the atomic ground level, as happens to all other muonic atoms generated by the stray muons. 
A mosaic of specialized detectors, covering the majority of the solid angle seen by the target, progressively records the flow of the X-rays. The injection into the target of the laser radiation starts the actual measurement once the $\mu H$ atoms have reached a thermal equilibrium with the surrounding gas. The multi-pass optical cavity that encloses the majority of the target’s gas volume increases the transition probability, as it will be discussed later.
When the appropriate wavelength inducing the transition is attained, the time distribution of the muonic oxygen X-rays will abruptly change, given the energy dependence of the muon transfer rate from hydrogen to oxygen $\lambda_{pO}(E)$~\cite{bib:adamczak2012}. The typical time of this chain of process to occur is of the order of few hundreds of nanoseconds.

The proposal of the FAMU experiment was presented in 2013 at the Program Advisory Committee (PAC) of the RIKEN laboratory, co-owner at that time of the RIKEN-RAL pulsed muon beam facility. After the approval of this first declination of the FAMU program, two subsequent competing proposals with the same physics target have been proposed and approved by other muon facilities~\cite{bib:antognini2015}~\cite{bib:sato2014}. 
Since 2013 the FAMU Collaboration devoted several fruitful data taking at the RIKEN-RAL facility, showing that the proposed method, the beam characteristics and the selected detection systems were all suited to perform the $\mu H$ {\em 1S-hfs} measurement, as described in a series of papers illustrating both technical and scientific results~\cite{bib:adamczak2016}~\cite{bib:vacchi2016}~\cite{bib:vacchi2017}~\cite{bib:mocchiutti2018}~\cite{bib:mocchiutti2019}.
In particular, the FAMU Collaboration was able to: confirm the suitability of the transfer method for the oxygen~\cite{bib:pizzolotto2021}; investigate different gas configurations to determine the best final setting (i.e. the type and quantity of high Z element in gas mixture and the values of pressure and temperature of the gas target); estimate the minimal duration of a run in the final configuration.

\section{The FAMU experimental setup}
\label{sec:setup}

\begin{figure*}[htpb]
\centering
\includegraphics[width=.8\linewidth]{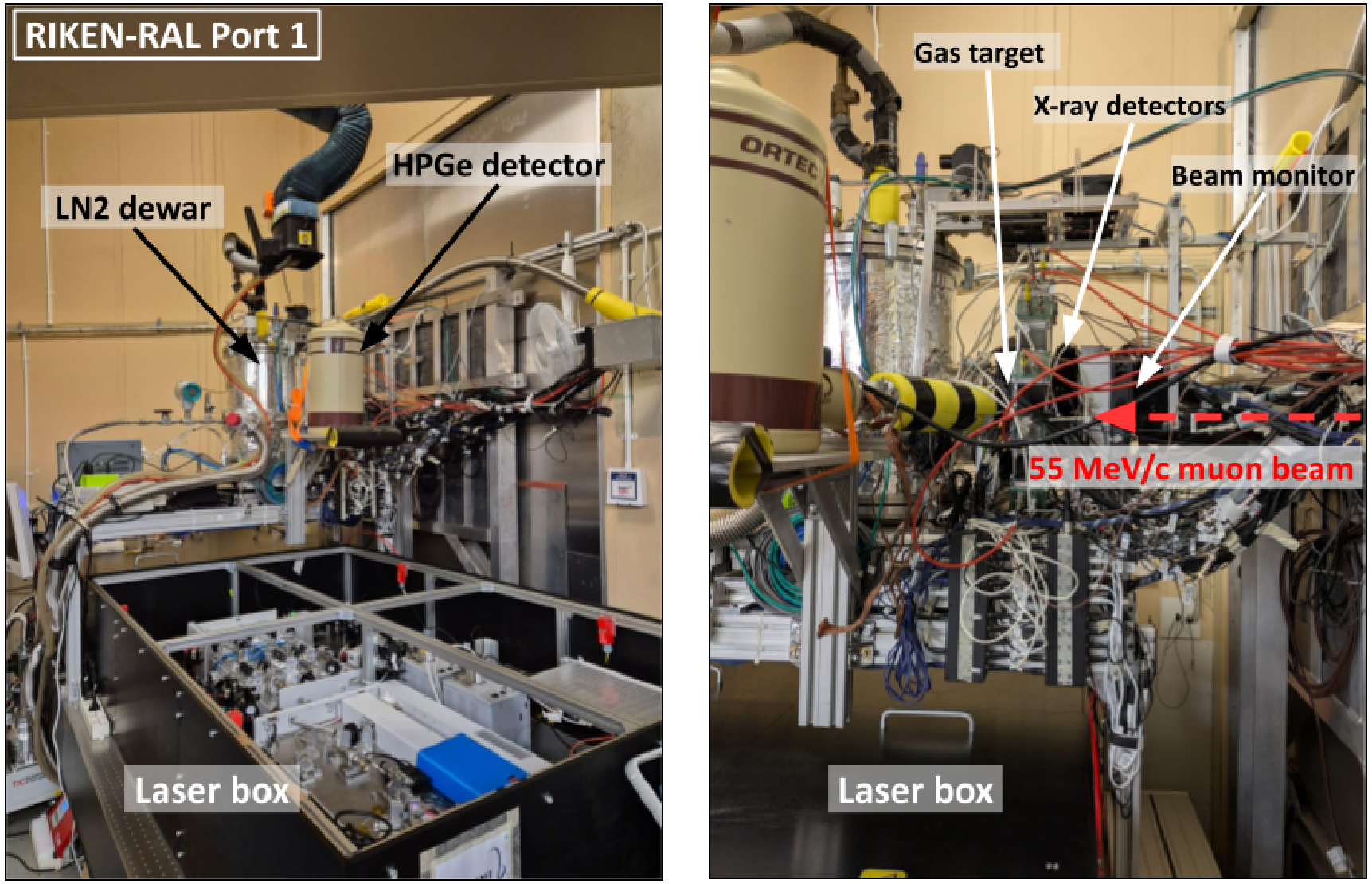}
\caption{Left: photograph of the fully mounted FAMU experiment at RIKEN-RAL Port 1, taken in September 2023 after the commissioning but before the first beam time. Right: zoom on the FAMU target and detectors.}
\label{fig:setup}       
\end{figure*}

The FAMU setup is basically made by: a high rate, tunable momentum, pulsed muon beam monitored by a custom beam monitor; a Hydrogen-Oxygen mixture gas target kept at cryogenic temperature; a mid-infrared laser system; an optical cavity to maximize the number of laser excitations inside the gas target; a system of X-ray detectors made by scintillating crystals read either by Photo-Multiplier Tubes (PMTs) or by Silicon Photo-Multipliers (SiPMs) arrays; a custom DAQ system allowing the event collection and storage. A photograph of the FAMU full setup installed in the Port 1 of the RIKEN-RAL muon facility, taken after the June 2023 beam commissioning, is shown in Fig.~\ref{fig:setup}.

\subsection{Muon beam characteristics and monitoring}
\label{subsec:beam}

\begin{figure*}[htpb]
\centering
\includegraphics[width=.6\linewidth]{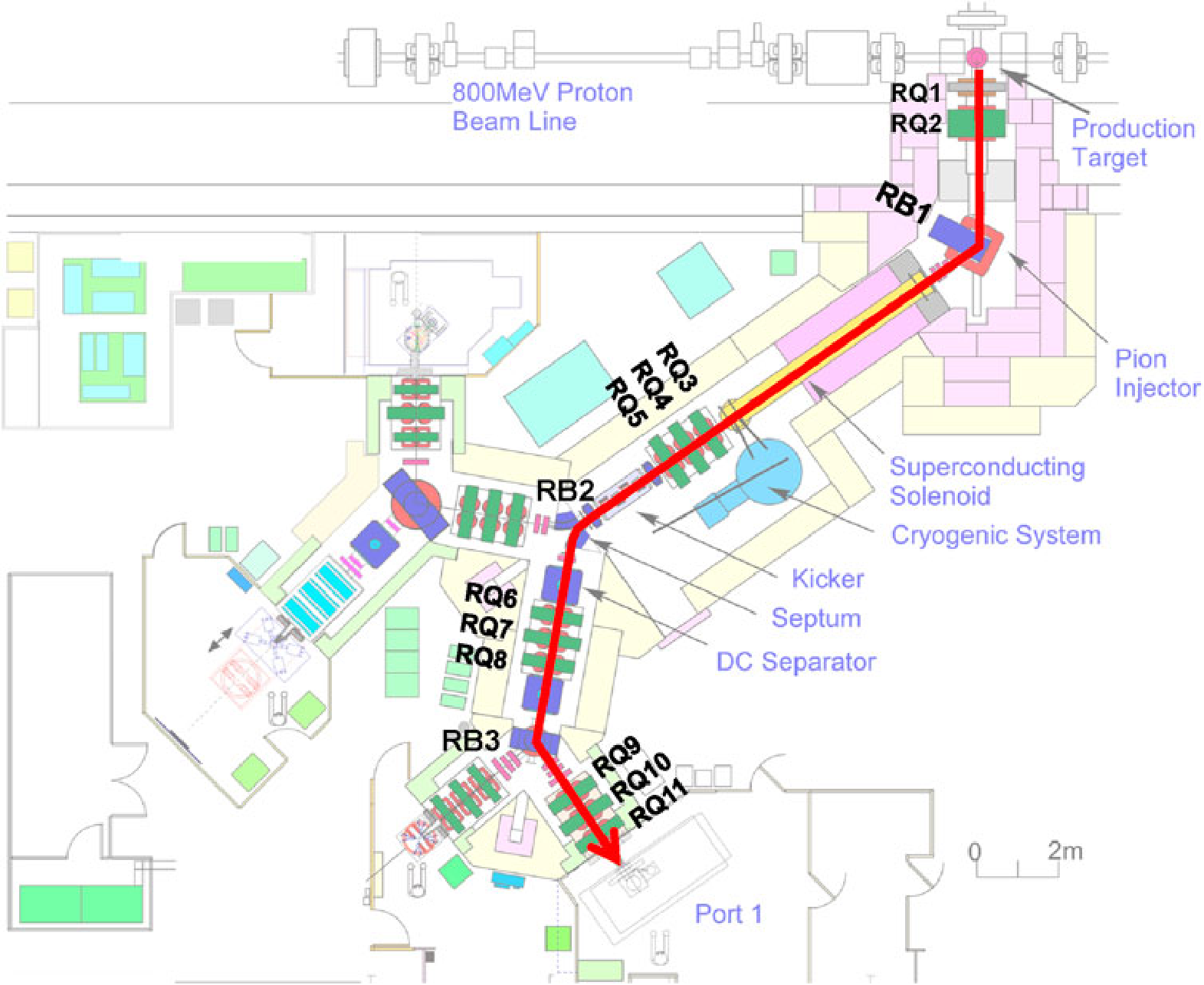}
\caption{Scheme of the path followed by muons directed to the FAMU target in the RIKEN-RAL muon facility. The positions of the bending (RB) and quadrupole (RQ) magnets involved in the beam delivery to FAMU are labelled.}
\label{fig:beam}       
\end{figure*}

The ISIS Neutron and Muon Source is currently the leading centre in Europe for the production of pulsed neutron and muon beams. The facility is located at the Harwell Campus in Didcot, Oxfordshire (UK), as part of the Rutherford Appleton Laboratory (RAL). 
The ISIS muon beam~\cite{bib:hillier2019}~\cite{bib:thomason2019} is produced by the collision of an 800 MeV proton beam with a target. The protons, accelerated inside the ISIS proton synchrotron, are extracted and directed to two different neutron production areas: Target Station 1 (TS1) and Target Station 2 (TS2). The proton beam extraction is carried out with a rate of 50 Hz, extracting particles from two proton bunches. For this reason, a proton spill is extracted every 20 ms, and it is composed of
two proton bunches separated by 320 ns. Every five proton spills, four are sent to TS1 and one to TS2. As a consequence, the average repetition rate of protons against the muon production target is 40 Hz. The pions produced in the muon production target by protons hitting a graphite target will decay with a time constant of $\sim$ 26~ns in muons and neutrinos, giving rise to the desired muon beam. 
Depending on whether the pions decay at rest inside the target or escape it, two species of muons can be extracted:

\begin{itemize}
    \item surface muons are $\mu^+$ originated from the decay at rest of $\pi^+$ produced close to the surface of the graphite target. The flux is generally high (over 10$^6$~muons/s), their momentum peaks at about 29~MeV/c, as they are produced at rest.
    \item  decay muons are $\mu^+$ or $\mu^-$ formed by transporting the pions out of the target, selecting their desired momentum, and letting them decay in order to produce a muon beam of a certain momentum. They are sent to the RIKEN-RAL muon facility, with a momentum-dependent flux (see Fig.~\ref{fig:beam}). RIKEN-RAL is capable of delivering muons between 17 and 120 MeV/c. In particular, the beam delivered to FAMU is a 55 MeV/c decay $\mu^-$ beam.
    \end{itemize}

\begin{figure*}[htpb]
\centering
\includegraphics[width=.4\linewidth]{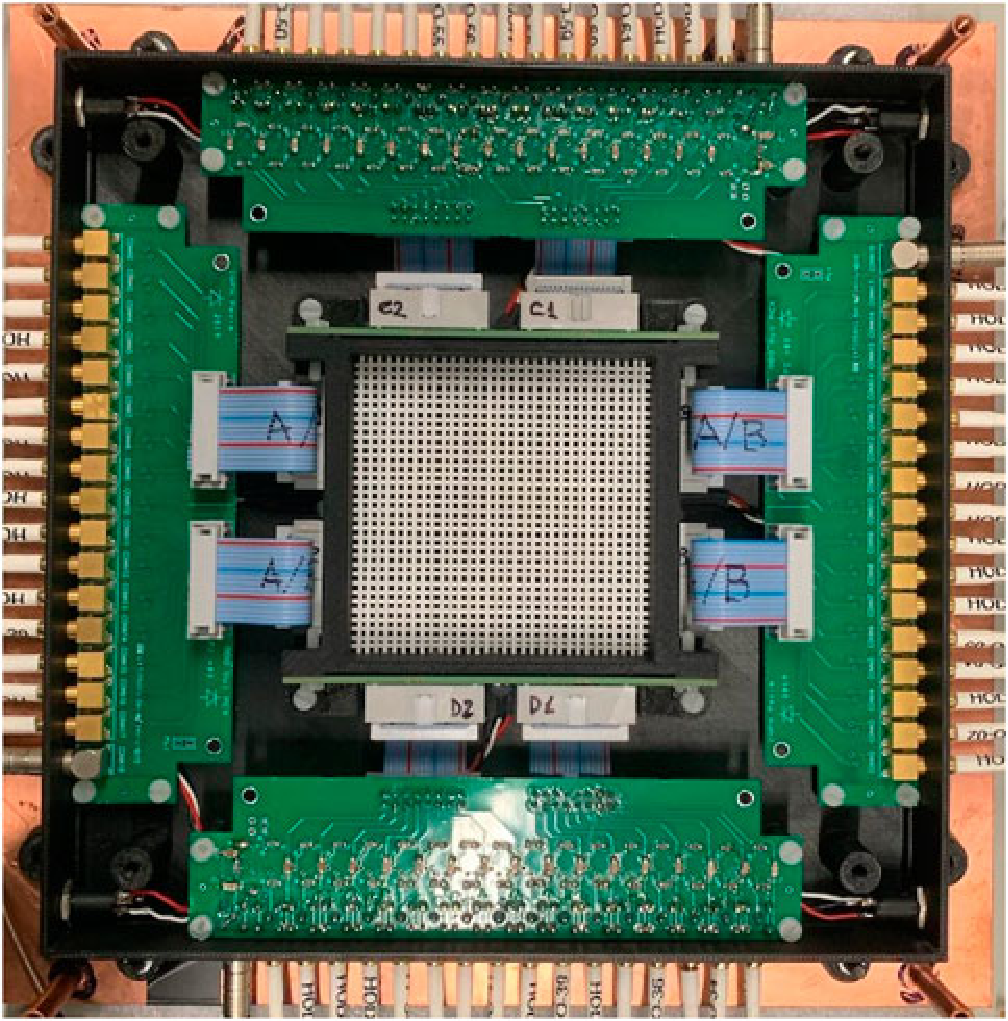}
\caption{Internal view of the FAMU beam monitor: the 32+32, 1~mm width scintillating fibers with TiO$_2$ EMA paint coating can be seen in the middle.}
\label{fig:hodo}       
\end{figure*}

The FAMU observable, consisting in the time distribution of the muon transfer events from hydrogen to oxygen as described in Sec.~\ref{sec:experiment},
is strongly dependent on the flux of the muon beam injected in the target, which changes the number of muonic hydrogen atoms formed. As a consequence, the presence of a beam monitor in FAMU is clearly crucial both for beam control and data normalisation. For this purpose, after a series of prototypes described in references \cite{bib:carbone2015}~\cite{bib:bonesini2017}~\cite{bib:bonesini2019}, a scintillating fibre-based hodoscope has been implemented as the muon beam monitor for the FAMU experiment. It was designed and built by INFN Milano Bicocca, INFN Pavia and INFN Roma3. It consists of two planes of polystyrene scintillating fibres (with perpendicular directions) read-out on one side by an Hamamatsu S12751-50P SiPM, with 50~$\mu m$ cells, see Fig.~\ref{fig:hodo}. The detector's fiducial area is 64~$\times$~64~mm$^2$. The 1 mm-pitch fibres are model BCF-12 by Bicron/Saint-Gobain/Luxium, with blue scintillation light (peak wavelength 435~nm), 3.2~ns of decay time, and a light yield of $\sim$~8000 photons/MeV for Minimum Ionising Particles (MIPs). The fibres, with white TiO$_2$ EMA (Extra Mural Absorber) coating, 10-15~$\mu m$ thick, to avoid cross-talk and interspaced by 1~mm from each other, are read-out on alternate ends by SiPMs soldered on four Printed Board Circuits (PCBs). Each PCB provides the same bias to all 16~SiPMs and fans out the 16~signals through MCX connectors. In the FAMU Data Acquisition System, hodoscope signals are fanned out through MCX connectors and acquired using two CAEN V1742 digitisers (32~channels, 5~GS/s, 12~bit). 

This detector, initially designed for beam focusing and centering, is now capable of measuring the muon beam flux. This has been made possible by measuring the total charge deposited in the detector fibres during a beam spill ($Q_{tot}$) and taking into account the detector geometry and the muon energy loss at fixed 55 MeV/c momentum. For this specific geometry, the flux can be calculated as $\varphi[s^{-1}] = k Q_{tot} [\text{ADC ch.}]$, where the constant $k$ is calculated as:
\begin{equation}
    k = \frac{r \ [\text{Hz}]}{(W_2 + \frac{W_1}{\eta}) \ Q_\mu\ [\text{ADC ch.}]}\text{.}
\end{equation}
Here, $r$ is the fixed average beam repetition rate (40~Hz), $Q_\mu$ is the average charge deposited by a single muon interacting with two fibres, determined experimentally through a dedicated measurement~\cite{bib:rossini2024}, and $(W_2 + W_1/\eta)$ is a simulated correction factor taking into account the fraction $W_j$ of muons interacting with $j$ = 1,2 fibres, and the ratio $\eta$ between the charge released in two fibres over that in one fibre. 

\begin{figure*}[htpb]
\centering
\includegraphics[width=.49\linewidth]{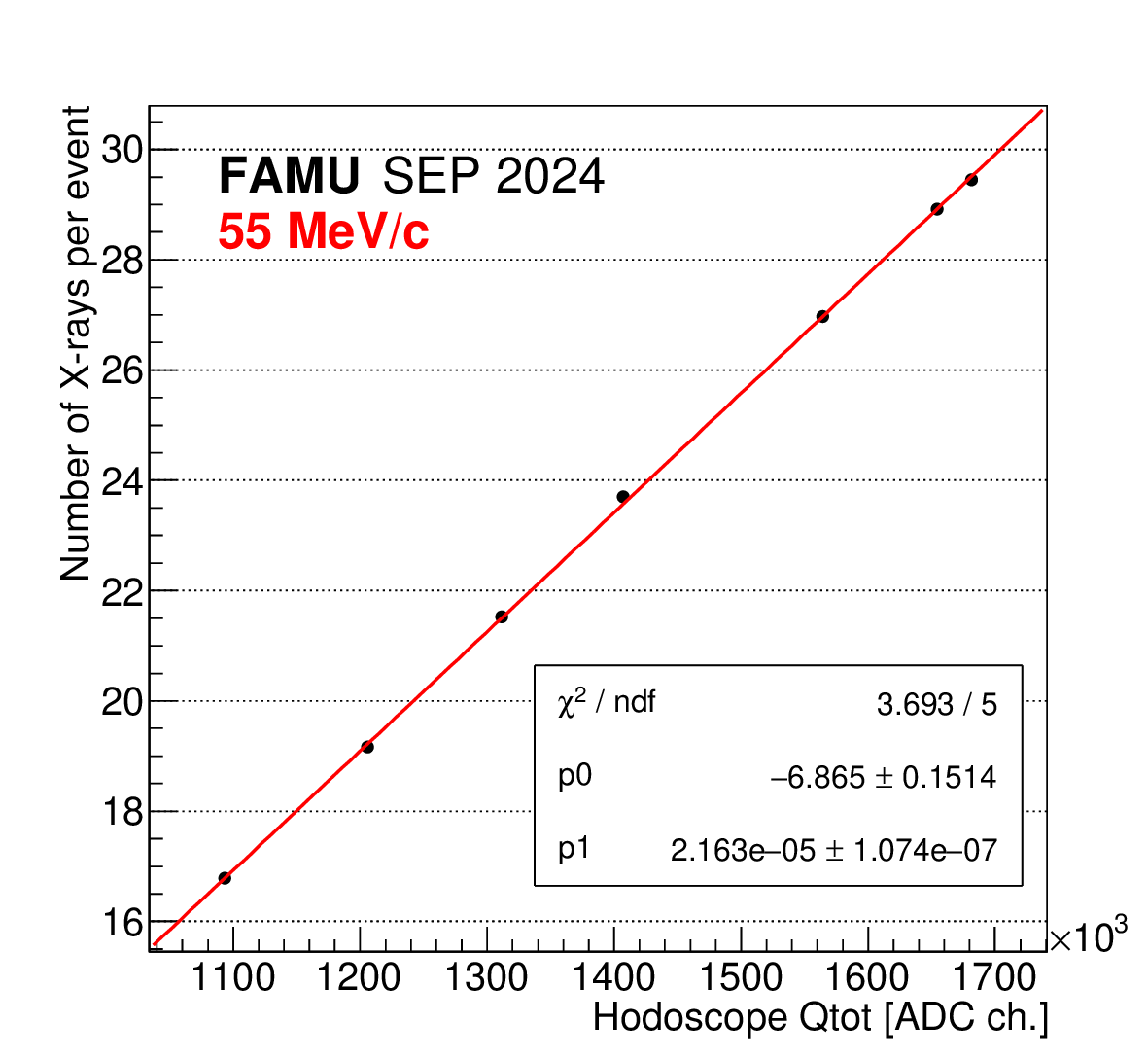}
\includegraphics[width=.49\linewidth]{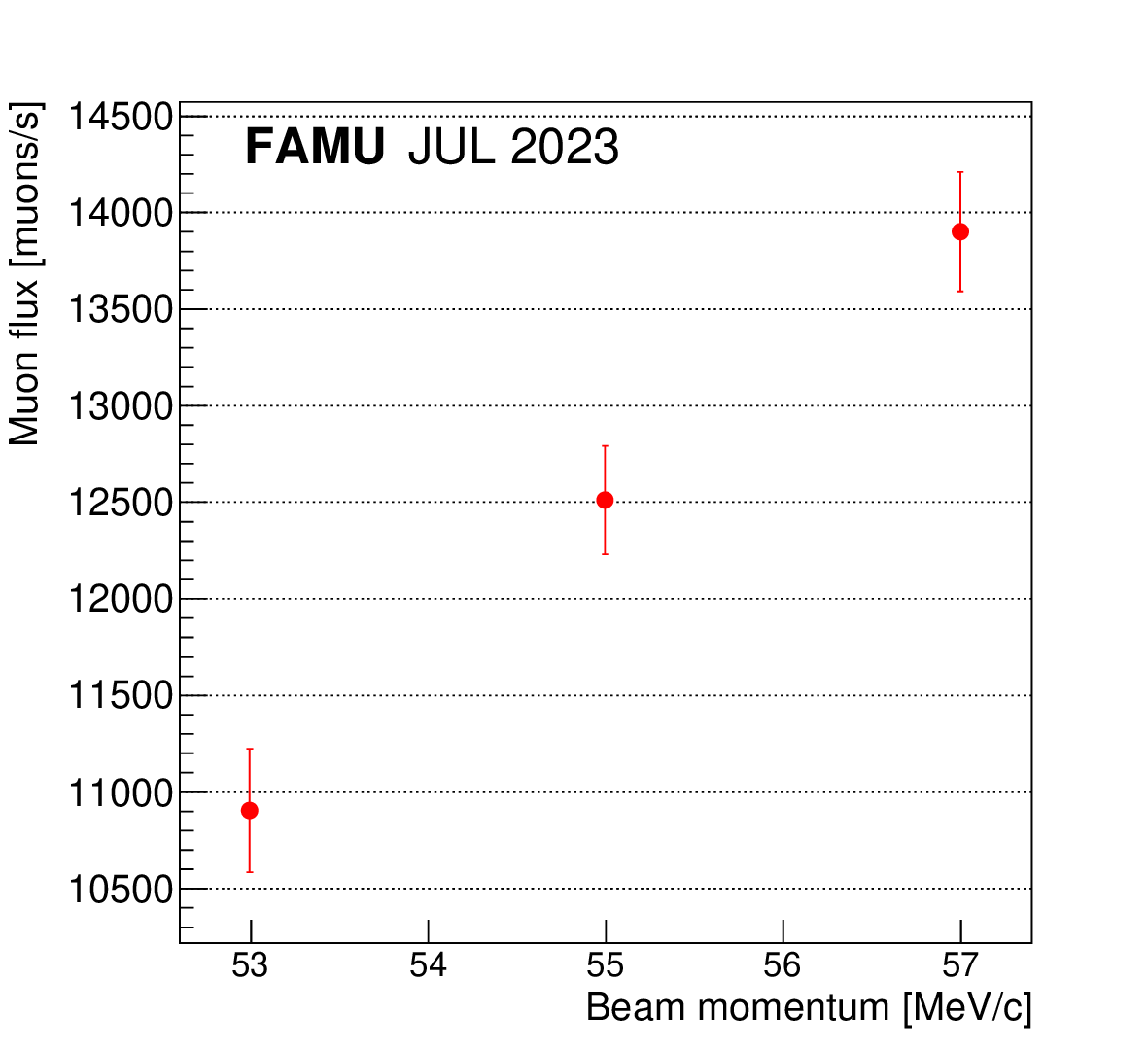}
\caption{Left: beam monitor flux linearity, tested by varying the flux through a dedicate de-tuning of bending magnet RB1, and comparing the charge deposited in the scintillating fibres ($Q_{tot}$) with the number of X-rays measured by the detectors, which is proportional to the number of muonic atoms in the target, hence to the muon flux. Right: measurement of the momentum-dependent muon flux around the FAMU working point, measured in the standard FAMU focusing condition.}
\label{fig:hodo:plots}       
\end{figure*}

The muon flux measurement enables on-line beam line monitoring and data normalisation as a function of the number of muons entering the gaseous target. Fig.~\ref{fig:hodo:plots} shows a linearity plot for the beam monitor, in which the value of $Q_{tot}$ is compared to the number of X-rays measured by the FAMU detector system, which is proportional to the number of muons on target, hence on the flux. The right plot shows a set of flux estimations as a function of momentum in the FAMU magnet setting.

\subsection{Target system and multi-pass optical cavity}
\label{subsec:target}

The kinetic transfer of the muon to oxygen acts as a background contribution to FAMU, as previously observed by FAMU in dedicated test beams at RAL. The experiment would ideally work at the lowest possible temperature for the hydrogen-oxygen mixture. However, the condensation temperature for O$_2$ limits the
temperature to be above $\sim$ 60~K. For this reason, it has been decided to use liquid nitrogen (77~K) as the target cooling medium. The target is filled with hydrogen with a percentage of oxygen of 1.5\% which, combined with a pressure of 7 bar, is the optimal choice for the 80~K target temperature~\cite{bib:pizzolotto2020}.

\begin{figure*}[htpb]
\centering
\includegraphics[width=.8\linewidth]{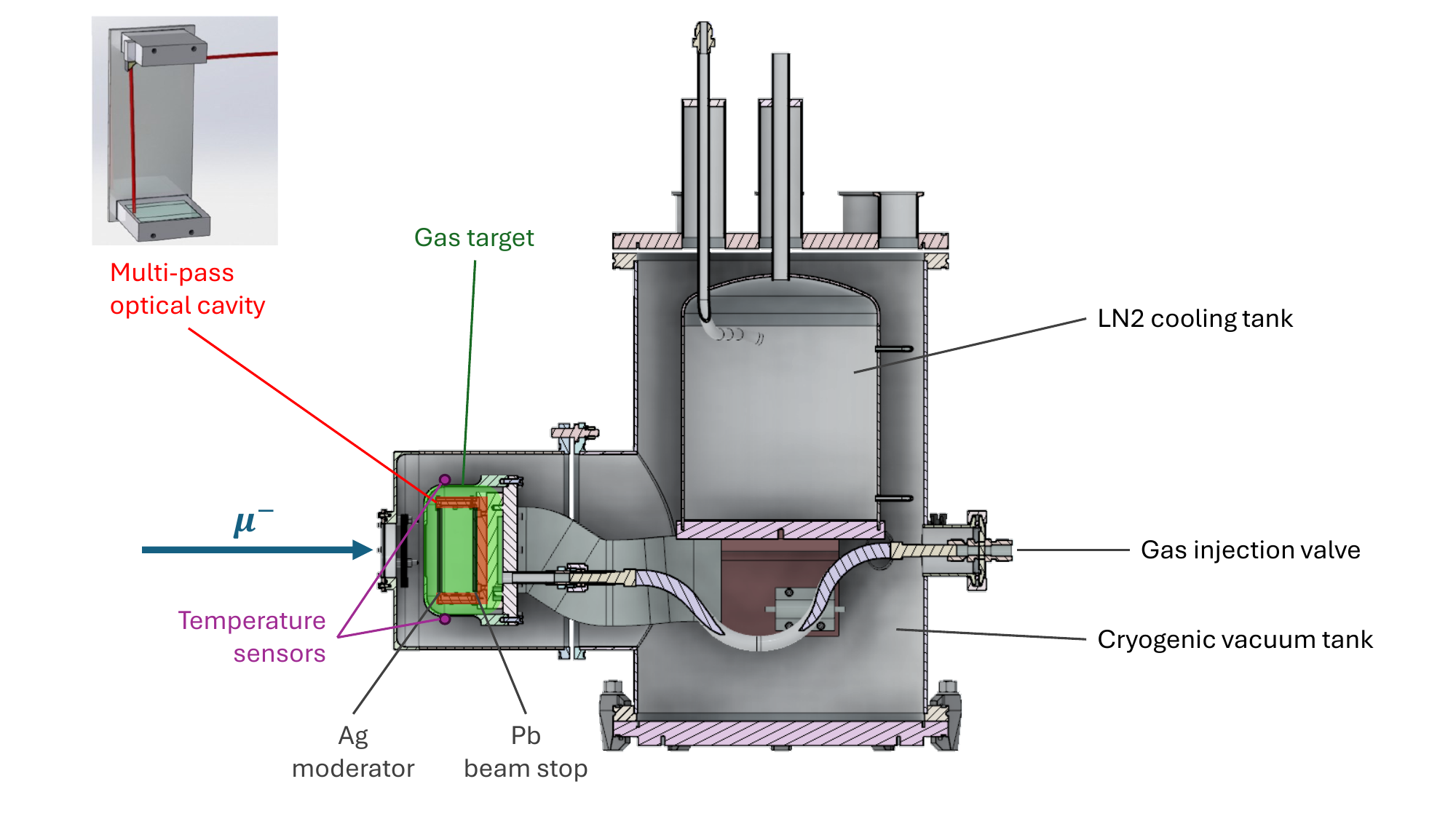}
\caption{Side section view of the target cryostat (CAD drawing), where the
main components are marked. The three detector-holding rings are attached to the target cryostat around the target chamber (green volume), as shown in Figure \ref{fig:detectors}.}
\label{fig:target}       
\end{figure*}

The target cryostat was designed to best satisfy all the FAMU requirement, namely: a thin high-Z muon beam entrance window, in order to maximise the momentum loss in the entrance window and therefore the muon capture in the gas; low-Z materials in other directions to let 100-200~keV X-rays out of the
target, where detectors are placed; an optical window, needed to inject the laser beam, and fitting an optical cavity to maximise the gas-laser interaction. The current target has been simulated, optimised, designed and mounted in collaboration with CRIOTEC Impianti Srl to match the requirements. A lateral section of the FAMU target cryostat is shown in Fig.~\ref{fig:target}. In
particular, the chamber containing the gas is shown in green and it is cooled by thermal contact with a tank containing 5 litres of liquid nitrogen at 77~K. This solution enables vibration-free cooling, which is crucial to avoid laser misalignments during the experimental runs. In order to monitor the gas condition, the target chamber is equipped with a pressure sensor (sensitivity 0.01~mbar) and two temperature sensors (sensitivity
1~mK). These sensors play an important role in assessing the tightness of the chamber and correcting for small gas leakage.

The muon beam exits the beam pipe, crosses the FAMU beam monitor and then passes through a beam collimator composed of lead
bricks. It is aimed at injecting the muon beam only in the central area of the target within the Multi-pass Optical Cavity (MOC). Muons enter the cryostat through a 0.2~mm aluminated Mylar window, then cross the 1~mm-thick aluminium wall of the gaseous target chamber, and lastly encounter a 0.6~mm layer of silver. The role of this Ag layer is to
maximise the number of muons stopping in the gas by slowing them down just before entering the volume illuminated by the laser. Each mirror of the MOC ~\cite{bib:pizzolotto2020} is constituted by three parts of Fused Silica coated with ZnS/Ge multi-layers to provide the best possible reflectivity $R$ = 99.890(2)\% at 6.78~$\mu$m. The three parts of the mirrors are designed in order to confine the light inside the cavity with a quasi-chaotic pattern. At this scope the ends of mirrors have a cylindrical shape while the central part has a flat surface. The MOC is aimed to enhance the transition probability by a factor of $1/(1-R)$, compared to Eq.~\ref{eq:probability}.
It is studied to let the laser beam bounce back and forth, minimising its power loss.
The alignment of the laser beam within the MOC is a very complex operation related to the fact that the MOC is closed inside the target that in operating condition is at T = 80~K and P = 7~bar.
Therefore an alignment procedure has been realized by using a ``twin" MOC placed outside the target as a reference.
The laser beam, from the optical table, is sent into the target by means of a lift (periscope) composed by two Off-Axis Parabolic Mirrors (OAPM). One of the OAPM is mounted on the table while the second one is in "L" shape aluminium mount directly connected to the target chamber. A beam splitter plate can be housed for alignment purposes between the OAPM\#2 and the window of the target chamber. 

\begin{figure*}[htpb]
\centering
\includegraphics[width=.7\linewidth]{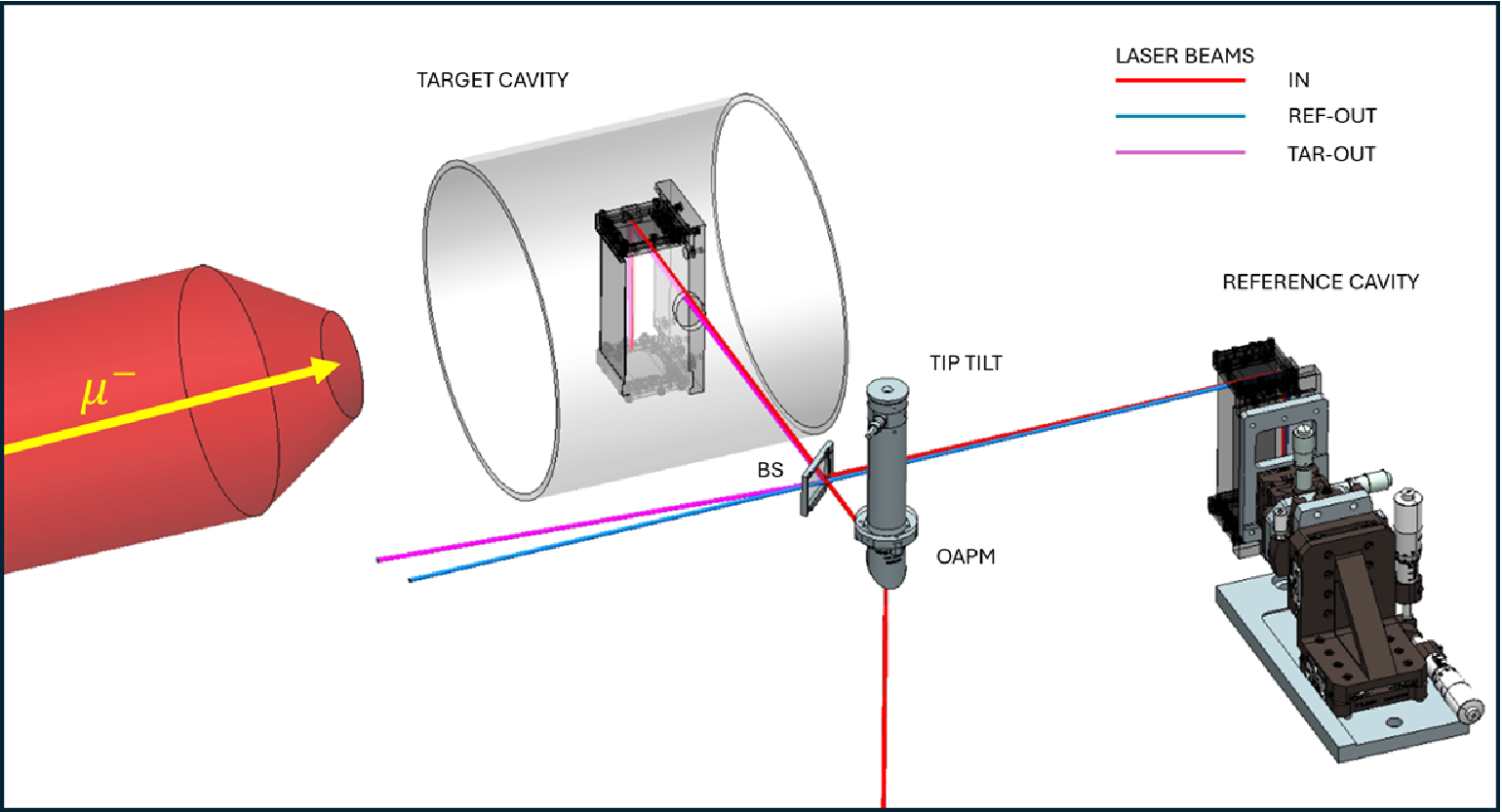}
\caption{Laser Injection Systems: sketch of the alignment procedure: a red laser beam, well overlapped with the 6.8~$\mu$m radiation, is directed towards the two identical Multi-pass Optical Cavity, namely the {\em reference cavity} and the {\em target cavity}. The injection is assured by aligning the reference cavity in a multi-pass optical configuration, being sure that the laser arrives at both cavities exactly the same way. BS stands for Beam Splitter and OAPM stands for Off-Axis Parabolic Mirror.}
\label{fig:moc1}       
\end{figure*}

In Fig.~\ref{fig:moc1} has been reported a sketch of the target ensemble with the external cavity on the side. The reference cavity is mounted on a roto-translation stage at the same height of the target cavity and it is placed on an aluminium support on the right side of the periscope.
The target cavity will be aligned by superimposing the reflected beam merging from the target cavity with the reflected beam exiting from the reference cavity with a visible laser beam at 632 nm.
The alignment procedure is carried out in two steps. First, we move the reference cavity with respect to a laser beam that hits orthogonally on the bottom mirror of the target cavity, in order to overlap the beams reflected from the two cavities. This ensures that both cavities are aligned to the same position relative to the laser beam. Then, we align the target cavity in a multi-pass configuration by monitoring the optical path in the reference cavity:

\begin{itemize}
    \item a visible laser (red) is injected in the target cavity and, by adjusting the position OAPMs, a configuration with a back reflection overlapped on the input one is obtained. In this way the laser beam impinges orthogonally on the bottom mirror of target cavity.
    \item A beam splitter at 45$^{\circ}$ is introduced after the OAPM \#2. The laser beam is split in two parts and sent to two cavities. The distance between the beam splitter and the two injection mirrors is the same.
    \item By means the translators and rotators and monitoring the reflected spots light emerging from the two cavities, the reference cavity is aligned in the same way of target cavity. This step guarantee that the optical path in reference cavity is the same as that in target cavity.
    \item By moving the OAPMs with motorized tip/tilt the number of reflections in the reference cavity is maximised, with warranty that the optical path in the target cavity will be very similar.
    \item The visible beam laser is switched off, the beam splitter is removed and the infrared laser is injected in the target cavity.
\end{itemize}

\subsection{Narrowband mid-infrared pulsed laser}
\label{subsec:laser}

\begin{figure*}[htpb]
\centering
\includegraphics[width=.7\linewidth]{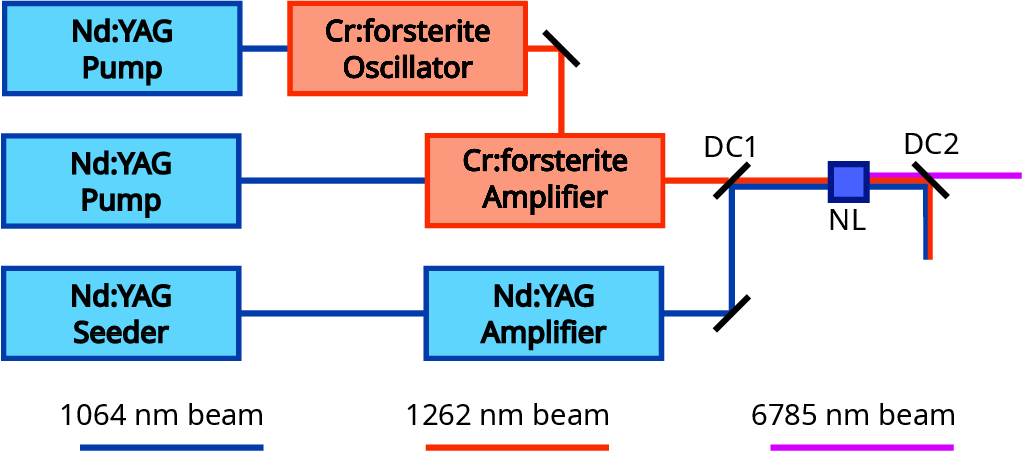}
\caption{Simplified sketch of the FAMU laser system representing the pump and signal formation and the coupling of the beams inside the non-linear (NL) crystal. DC1 and DC2 are dichroic mirror used respectively to superimpose the beams and to isolate the 6.78~$\mu$m.}
\label{fig:laser}       
\end{figure*}

The FAMU laser system, extensively described in~\cite{bib:baruzzo2024}, is a custom-made Mid Infra-Red (MIR) laser system, whose main characteristics are: wavelength around 6789~$\pm$~3~nm, energy over 1~mJ, linewidth below 0.07~nm, tunability step below 0.03~nm, pulse duration below 10~ns and 25~Hz repetition rate. The FAMU laser performs excellently and exceeds the goal requirements in terms of energy, linewidth and tunability step. The rate is half of the synchrotron rate in order to inject the laser every other beam
spill, and use the spills with no laser as a background measurement. As of 2024 and at the best of our knowledge, the FAMU laser system is the most powerful pulsed 6.8~$\mu$m source available, combining narrow linewidth and fine tunability, and capable of delivering over 1.5~mJ of energy. 
The setup consists of two lasers: a fixed 1064~nm Nd:YAG Master-Oscillator-Power-Amplifier (MOPA), and a tunable (1262~$\pm$~5) nm Cr:forsterite oscillator, which is then further injected in an 16-passes 3-stages Cr:forsterite amplifier. The Cr:forsterite oscillator is featured with a diffraction grating, and the two cavity mirrors are placed on piezoelectric motors in order to allow the adjustment of the wavelength. The two beams are then properly coupled and injected in a non-linear Difference Frequency Generator (DFG), in this case a crystal of barium-gallium selenide (BaGa$_4$Se$_7$). The output of the crystal is a laser beam with frequency given by the relation $\lambda^{-1}_{DFG}$ = $\lambda^{-1}_{Nd:YAG}$ - $\lambda^{-1}_{Cr:forst}$, which is in the required wavelength range. By tuning the wavelength of the Cr:forsterite laser it is therefore possible to tune the MIR wavelength for the experiment. The entire system can be monitored remotely via dedicated software, specifically designed to interface with a set of sensors. Based on their feedback, the system can be adjusted either automatically or manually using piezo motors. 
A schematic layout of the FAMU laser system is shown in Fig.~\ref{fig:laser}.
The laser was designed and perfected by INFN Trieste in collaboration with Elettra Sincrotrone in Trieste (Italy) and then installed, characterised extensively and further improved at RAL in collaboration with Università della Campania/INFN Napoli~\cite{bib:stoychev2014}~\cite{bib:stoychev2015}~\cite{bib:stoychev2019}~\cite{bib:stoychev2020}. 

\subsection{X-ray detectors}
\label{subsec:labr}

The X-ray detection system of the FAMU experiment  requires: 
\begin{itemize}
    \item a good energy resolution at low energy ($\sim 130-150$~keV) to detect the signal muonic oxygen lines;
    \item a short signal fall time (less than 200-300~ns) to separate the delayed muonic X-rays (signal) from the prompt background and to reduce the signal pile-up;
    \item a large solid angle coverage to increase statistics.
\end{itemize}

\begin{figure}[htbp]
\centering
\includegraphics[width=.6\linewidth]{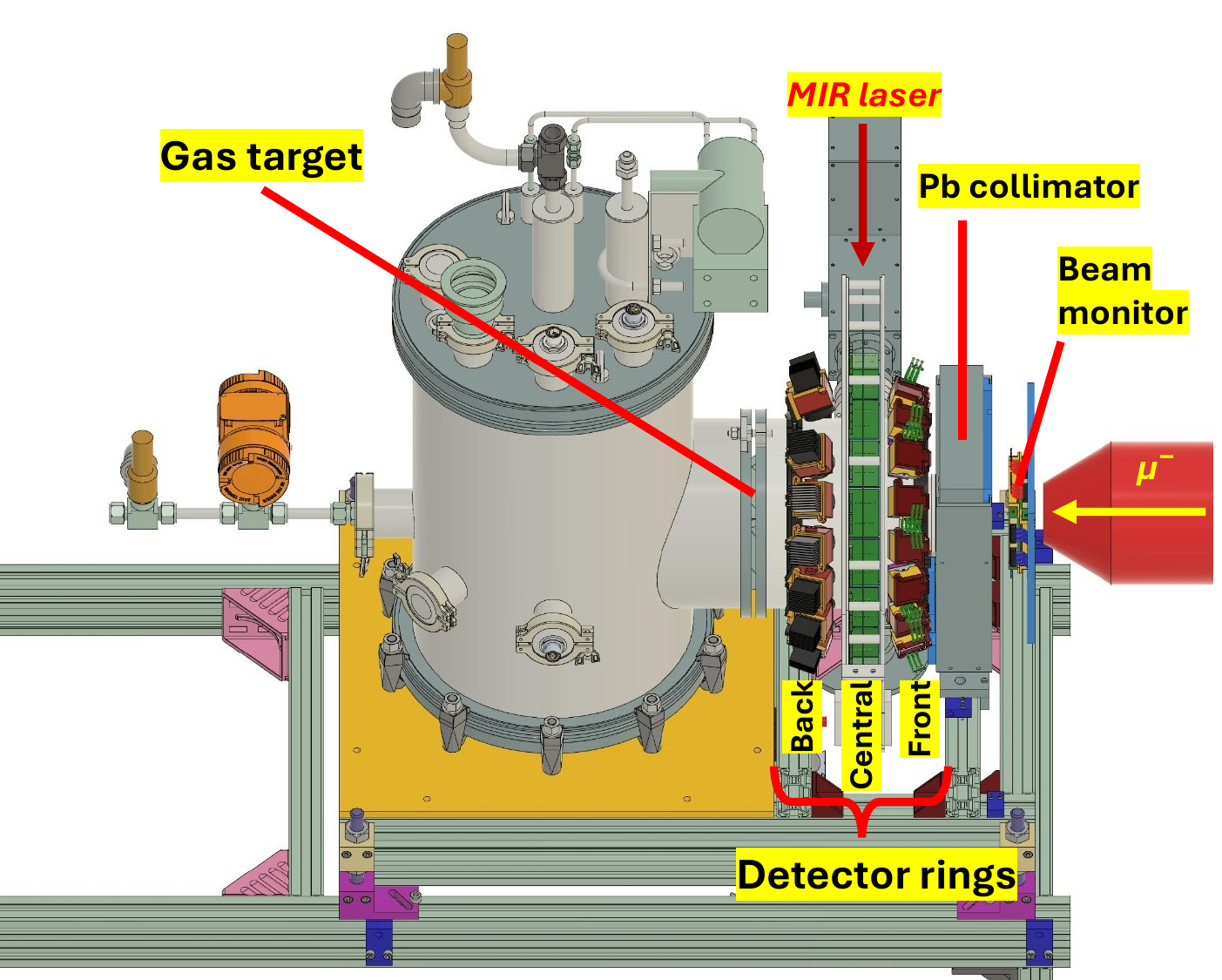}
\caption{Position of the three detector rings in the FAMU setup. The HPGe detector is omitted for improved illustration clarity.}
\label{fig:detectors}
\end{figure}

The FAMU experiment makes use of different types of crystal detectors, arranged in three rings around the target chamber along the beam line and pointing to the centre of the optical cavity. All these detectors are based on lanthanum bromide doped with cerium scintillating crystals (LaBr$_3$:Ce) and have different readout system. Detectors with PMT readout were developed at INFN Bologna, while the ones with SiPM array readout were conceived and built at INFN Milano Bicocca, in collaboration with INFN Pavia. The FAMU setup for the 2023 data taking (so called {\em setup A}) is based on 34 such detectors positioned around the target as shown in Fig.~\ref{fig:detectors}:

\begin{itemize}
\item six 1" diameter, 1" thick cylindrical crystals read by Hamamatsu R11260-200 PMTs with a custom high voltage divider ({\em ``LaBr"})~\cite{bib:baldazzi2017};
\item sixteen 1" diameter, 1/2" thick cylindrical crystals read by Hamamatsu S14161-6050-04-AS SiPM arrays ({\em ``MIB045-MIB063"}) \cite{bib:bonesini2022} \cite{bib:bonesini2023};
\item twelve 1/2" cubic detectors read by Hamamatsu S13361-3050-04-AS SiPM arrays ({\em ``MIB071-MIB099"})~\cite{bib:bonesini2020}~\cite{bib:bonesini2021}.
\end{itemize}

The six detectors with PMT readout are placed in the central ring, together with six 1" detectors with SiPM array readout. The remaining ten 1" detectors with SiPM array readout are placed in the upstream ring, while the cubic 1/2" ones are positioned in the downstream ring~\cite{bib:rossini2024_1}. 

In the 2024 data taking the twelve 1/2" cubic LaBr$_3$:Ce {\em MIB} detectors were replaced by 1" diameter, 1/2" thick cylindrical detectors with SiPM array readout, increasing the solid angle coverage by $\sim$~30~\% in one of the regions with highest statistics of delayed oxygen X-rays (so called {\em setup B})~\cite{bib:rossini2024_2}.
The detectors with SiPM array readout have a better FWHM energy
resolution, as compared to the ones with PMT readout (see Tab.~\ref{table3}), at the cost of a slower fall time. This last feature has prompted a long R\&D to reduce detectors' fall time. 
While the 1/2" detectors with a SiPM readout made use of a conventional parallel ganging, the 1" detectors use an innovative 1-4 circuit from Nuclear Instruments srl, that reduces the fall time by a factor two, as shown in~\cite{bib:bonesini2022}~\cite{bib:bonesini2023_1}.
In this layout, the 1" SiPM array is divided into four sub-array, by using four nearby $6 \times 6~mm^{2}$ SiPM cells . 
Output signals from these are summed together and then amplified with a pole-zero compensation. In this way, the capacitances are reduced. At the end, the four sub-array signals are summed together and inverted to produce a positive output. To handle the temperature increase ($\sim 5\div7 ^{\circ}C$) due to the used TI 695 op-Amplifier ($\sim 1$W) a heat power dissipater had to be put in thermal contact with the backside of the used PCBs and an ad-hoc ventilation system had to be introduced in Port 1.

As explained in reference~\cite{bib:otte2017}, the SiPM gain drifts as a function of a varying temperature. To keep it stable a custom system, based on CAEN A7585 power supplies with temperature feedback, was studied, as reported in reference~\cite{bib:bonesini2022_1}. As the temperature in Port 1 is kept stable within $\sim \pm 1 ^{\circ}C$ and the temperature coefficient of Hamamatsu SiPM arrays is quite small ($ -34 $~mV/C), up to now, this online correction system has not been used.
Possible residual gain drift of detectors with a SiPM readout, during extended data taking periods, have been corrected offline by calibrations with the beam data themselves. At the end of this procedure, initial variations at the few per-cent level have been reduced to zero. Further details on the performance in the
beam for these detectors are reported in reference~\cite{bib:bonesini2025}.

For inter-calibrations, a commercial ORTEC GEM-S5020P4 High Purity Germanium (HPGe) detector was also installed. 
The timing and FWHM energy resolution of the three types of detectors are shown
in Tab.~\ref{table3}. FWHM energy resolutions, for detectors with SiPM readout, at $^{137}$Cs and $^{57}$Co peaks are from laboratory measurements at INFN Milano Bicocca, while at 141~keV (muonic silver peak) are from beam data at RAL with 55~MeV/c impinging muons.
For comparison, the FWHM energy resolution at the 141~keV muonic silver peak is ($1.26 \pm 0.17) \%$ from the HPGe detector, at the cost of a much longer fall time.
10-90$\%$ rise time and fall time are measured at the $^{137}$Cs peak. 

\begin{table}[htb]
        \label{table3}
\smallskip
\centering
\begin{tabular}{|l|c|c|c|c|c|}
\hline
             &  rise time (ns) & fall time (ns)  & R($\%$) @ $^{137}$Cs   &
 R($\%$) @ $^{57}$Co & R($\%$) @ 141 keV  \\  \hline
1" - PMT     & 14 $\pm$ 1 & $\sim$ 60  & 3.5-4.6   &  7.2-8.1  & 12.3 $\pm$ 1.2 \\
1" - SiPM    & 29.3 $\pm$ 1.5 & 147.1 $\pm$ 12.8 & 2.94 $\pm$ 0.14 &
               8.03 $\pm$ 0.39 & 7.0 $\pm$ 0.3 \\
1/2" - SiPM  & 42.8 $\pm$ 4.7  & 372.4 $\pm$ 17.4 & 3.27 $\pm$ 0.11 &
               8.44 $\pm$ 0.63 & 7.5 $\pm$ 0.3 \\
\hline
\end{tabular}
\caption{Average detectors performances in terms of rise and fall times, and energy resolutions for the full set of installed detectors. Sample averages and RMS deviations are reported.}
\end{table}

\subsection{Trigger and Data Acquisition}
\label{subsec:daq}

\begin{figure}[htbp]
\centering
\includegraphics[width=.7\linewidth]{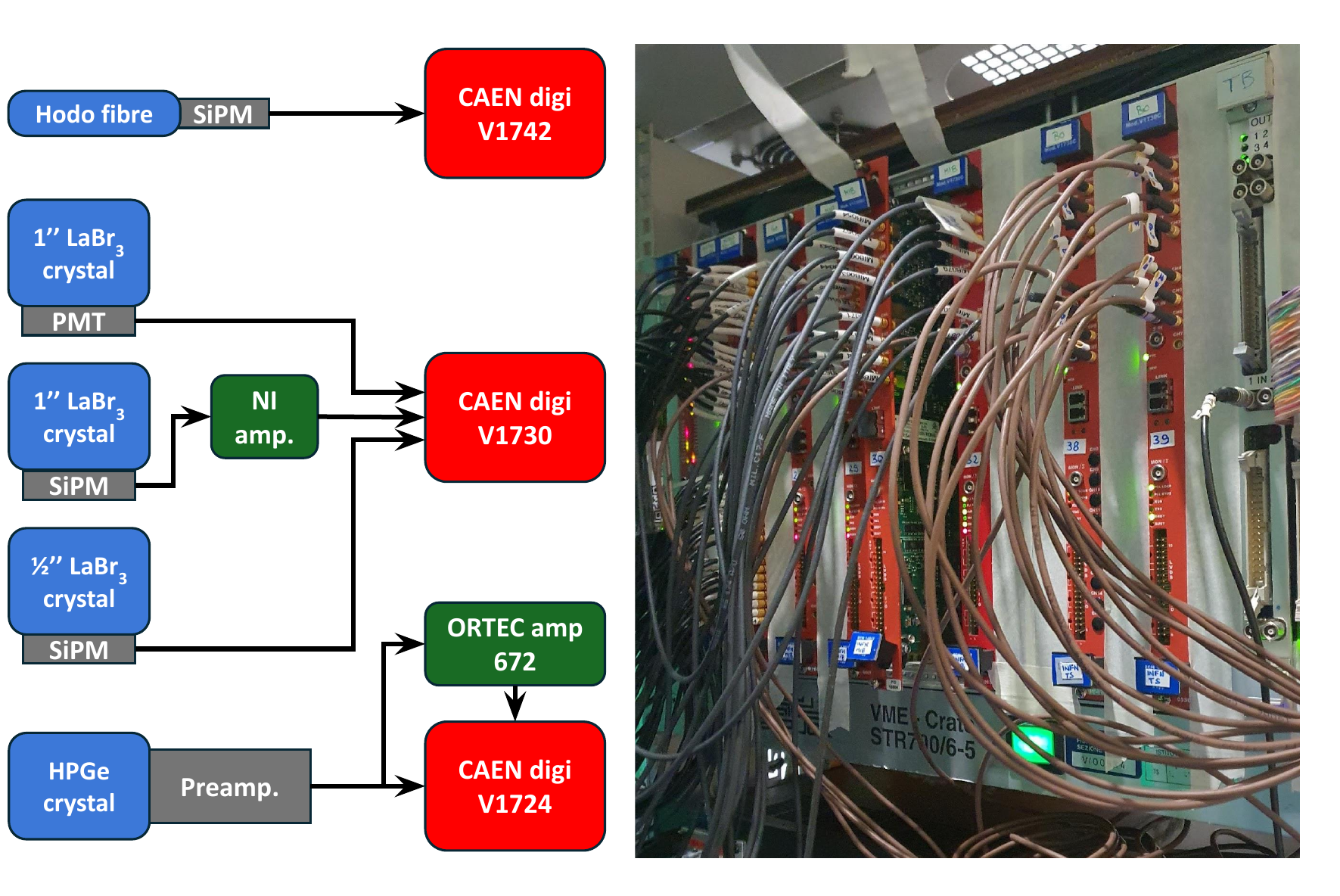}
\caption{Scheme of the DAQ electronics for all types of detectors (beam monitor fibres, scintillating crystals with different readout and HPGe detector) and close-up view of the main VME crate hosting the digitisers.}
\label{fig:daq}
\end{figure}

The data acquisition system of FAMU (FAMU-DAQ)~\cite{bib:soldani2019} handles six digitisers in two VME crates outside Port 1, to digitise all the required signals. Fig.~\ref{fig:daq} shows a scheme of the readout system for each type of detector in the FAMU setup, together with a picture of the main VME crate. The whole DAQ system is triggered by the muon beam and specifically by the Cherenkov scintillators placed in a hollow of the pion injection system. The beam hodoscope signals are fanned out from the detector with 64 MCX connectors and each of them is digitised as-it-is. Two CAEN V1742 digitisers  (32~channels, 5~GS/s, 12~bit) are used for this purpose, limited at 1~GS/s which is enough for the purpose.
All the signals coming from the scintillators, regardless of their readout system (PMT, SiPM Array + 4-1 circuit), are fanned out through MCX connectors and digitised using a total of six CAEN V1730/V1730S ADCs (8 channels,v500 MS/s, 14 bit).
The HPGe detector long signals are acquired with a CAEN V1724 (8 channels, 100 MS/s, 14 bit) in order to digitise over a longer time window and lower sampling rate. In particular, the signal is digitised both as-it-is and with
a shaping carried out with an ORTEC 672 amplifier with 3~$\mu$s shaping time.

The trigger system allows three modes of operation: random trigger for
hardware debug, external NIM trigger for normal beam operation and external TTL trigger for calibration. In particular, the latter is used to calibrate detectors with radioactive sources: one channel is used as a source of the
TTL trigger signal, which is then distributed to all digitisers through a trigger
board. During normal beam operation, instead, a NIM signal coming from the Cherenkov detectors in the pion beam line is distributed to the digitisers by the same trigger board.
The FAMU data acquisition system saves data in raw ROOT files, also called raw files. The software writes a file every 400 events, and 500 files (i.e. 200k events) form a run.
At this stage, data processing is carried out on-line through a custom program which runs on all raw files producing the so-called second-level files which are used for the high-level analysis.

The Cherenkov trigger used for the X-ray detector is not suitable for the laser, that needs a longer time to respond to generate light for illuminating the muonic hydrogen. The three laser sources that produce the 6.78~$\mu$m light each require two triggers, one to excite the active medium and the other to activate the Pockels Cell (PC), which initiates the laser pulse. The first trigger, commonly known as the Lamp trigger, must be sent to the laser system 140–150~$\mu$s before the laser pulse, while only 500~ns elapses between the PC trigger and the laser shot.
The Cherenkov trigger arrives 500~ns after the muon’s arrival, when the muonic hydrogen has already formed, making it incompatible with the laser system’s trigger-to-shot timing.

\begin{figure}[htbp]
\centering
\includegraphics[width=.7\linewidth]{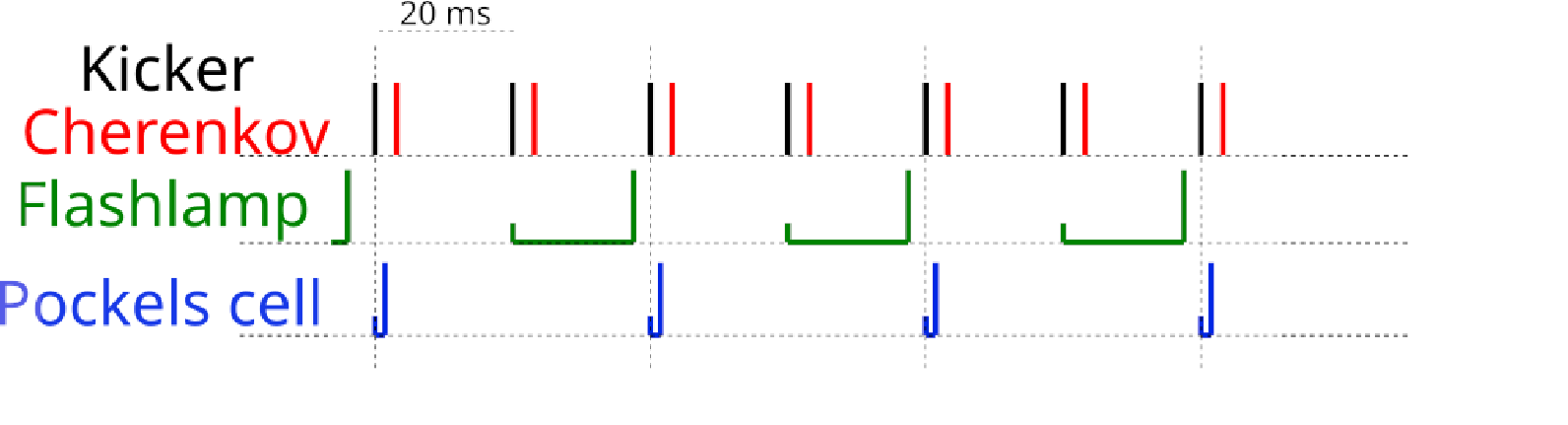}
\caption{The diagram illustrates the working principle of the laser trigger. Since the Cherenkov trigger arrives after the muons, it cannot be used to activate the laser. Instead, the solution is to use the kicker trigger: the PC is triggered by a signal that arrives 3.5~$\mu$s before the muons, while the flashlamps are activated by the previous signal.}
\label{fig:laser_trigger}
\end{figure}

Instead, the laser system uses a signal directly from the ISIS extractor, the device responsible for sending protons to the target stations. This signal reaches Port 1 approximately 3.5~$\mu$s before the muons arrive, providing enough time to serve as the trigger source for the PC. The previous extractor trigger is used for the lamp trigger. A scheme of the laser trigger is shown in Fig.~\ref{fig:laser_trigger}. Another characteristic of the laser is that it operates at 25~Hz, rather than 50~Hz like ISIS, requiring it to fire once every two muon pulses. The laser data required by the experiment, energy and wavelength measured shot-to-shot, are collected with a different computer respect the X-rays. This laser events are triggered by the trigger board through a microcontroller that is connected to the computer via a serial connection. The microcontroller also collects the event number from the trigger board, enabling a one-to-one matching of the laser and X-ray DAQ events.

\section{First operations at Port 1}
\label{sec:operations}

The first beam time assigned to the FAMU experiment by the scientific committees of the ISIS Neutron and Muon Source in July 2023 was used as a run for the beam, target, laser and commissioning of the full setup. Therefore, no physics data acquisition was made. After the FAMU successful commissioning phase at Port 1, the run for physics started with two beam times in October and December 2023, followed by two further beam times in 2024, for a total of 49 days of operations. Each data taking period was divided in batches, where a batch is defined as the set of contiguous measurements taken with the same laser wavelength: a batch number {\em ``Batch N"} is then used to uniquely identify each data sample within the same data taking period.

\subsection{FAMU commissioning run}
\label{subsec:commissioning}

The FAMU setup installation began already in 2019, but forced stops and delays were caused first by the COVID-19 pandemic and then by the ISIS facility long-shutdown for the TS1 project, completed in 2022. FAMU installation was completed by mid 2023, and the commissioning run was held from July 17th to July 23rd, 2023. The first operation consisted of the gas target emptying, filling and cooling with the H$_2$/O$_2$ mixture. 

\begin{figure*}[htpb]
\centering
\includegraphics[width=.45\linewidth]{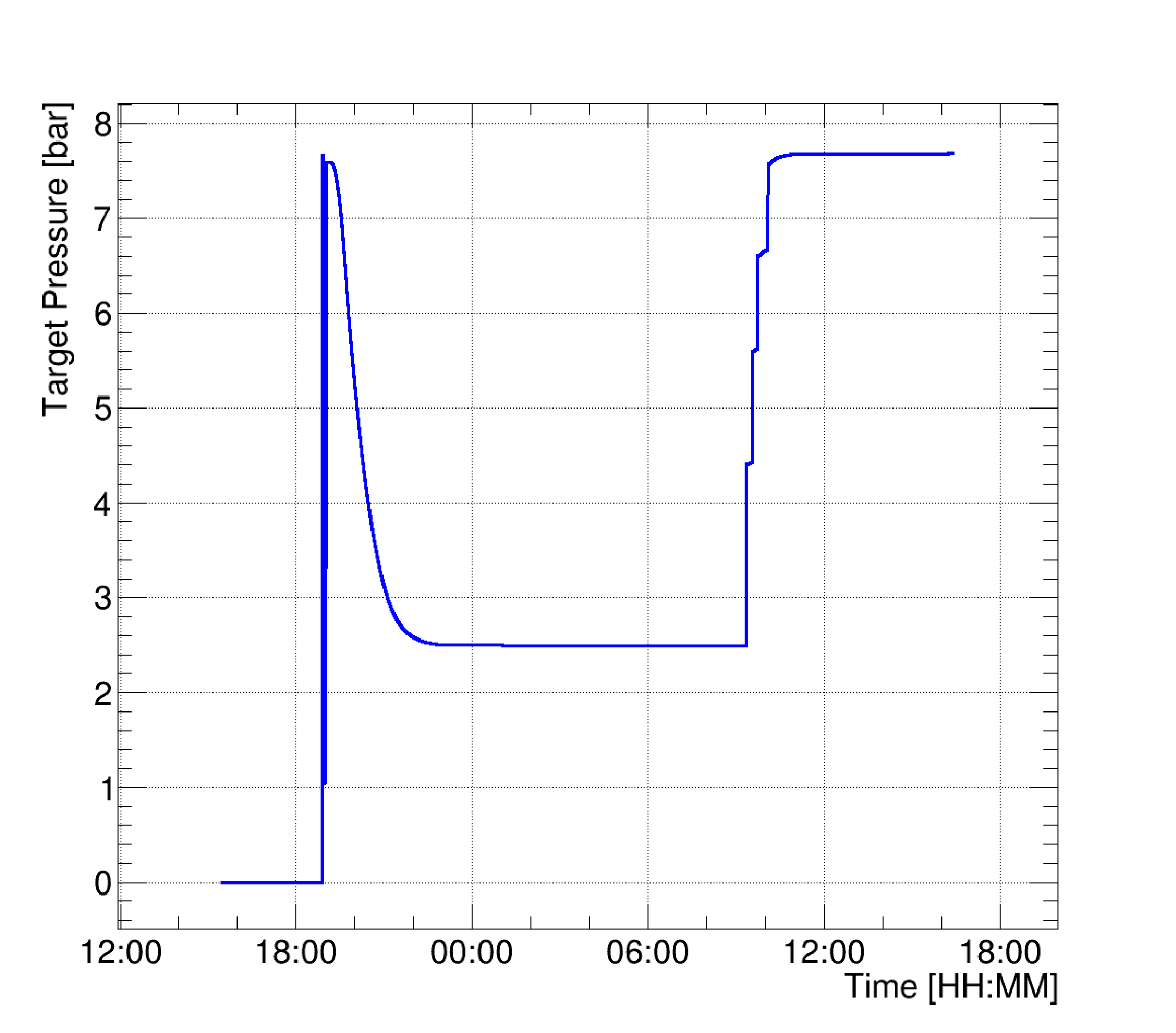}
\includegraphics[width=.45\linewidth]{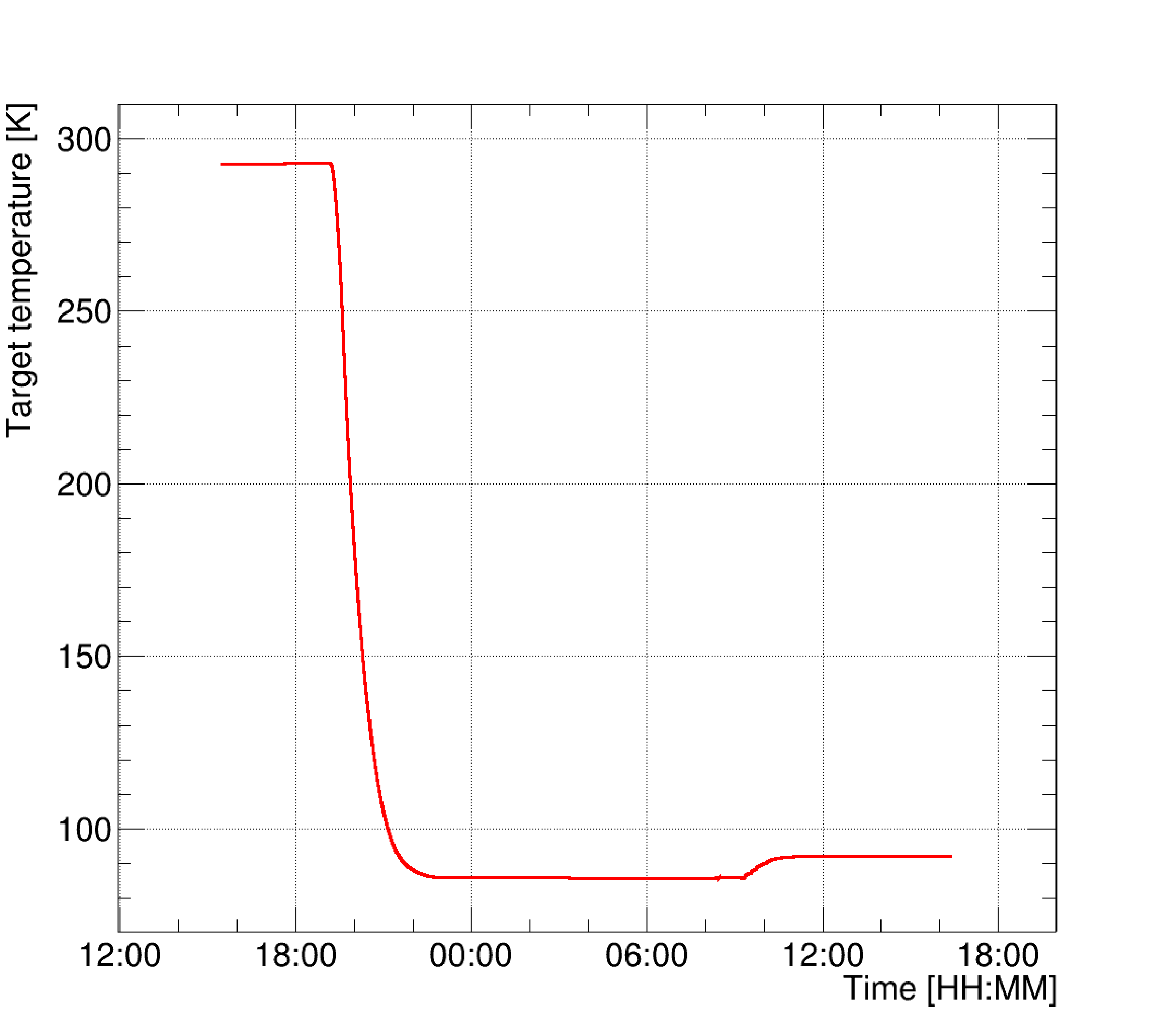}
\caption{Measured target pressure (left) and target temperature (right) as function of time during target filling and cooling.}
\label{fig:target_comm}       
\end{figure*}

The procedure is illustrated in Fig.~\ref{fig:target_comm}. The left panel shows the pressure measured in the target as a function of time. Before data collection, the target is evacuated to remove any residual gas. Then, it is cleansed by filling it twice with the gas mixture, followed by evacuation each time (the corresponding lines cannot be distinguished in the figure). Finally, the target is filled with the gas mixture to approximately 7.5~bar, and the cooling process begins by filling the apparatus tank with liquid nitrogen. The right panel of Fig.~\ref{fig:target_comm} displays the target temperature. As seen in the figure, the temperature drops from room temperature (approximately 290~K) to about 85~K within a couple of hours. Consequently, the pressure in the target decreases to around 2.5~bar. The system stabilizes within a few hours. In this example, we allowed it to stabilize overnight, though a couple of hours would have been sufficient. After the stabilization period, the target is slowly refilled to 7.5~bar using a stepwise procedure (as shown in the left panel of Fig.~\ref{fig:target_comm}) to verify the optical cavity alignment. As the gas quantity in the target increases, thermal dispersion through the optical window also rises, causing the target temperature to increase from 88 K to approximately 91~K. This temperature provides a sufficient gap relative to the de-excitation kinetic energy to observe the expected signal.
The target system remains stable throughout the entire data taking period, with temperature fluctuations of less than one degree and a slow pressure decrease of less than 2\% over two weeks.

\begin{figure*}[htpb]
\centering
\includegraphics[width=.45\linewidth]{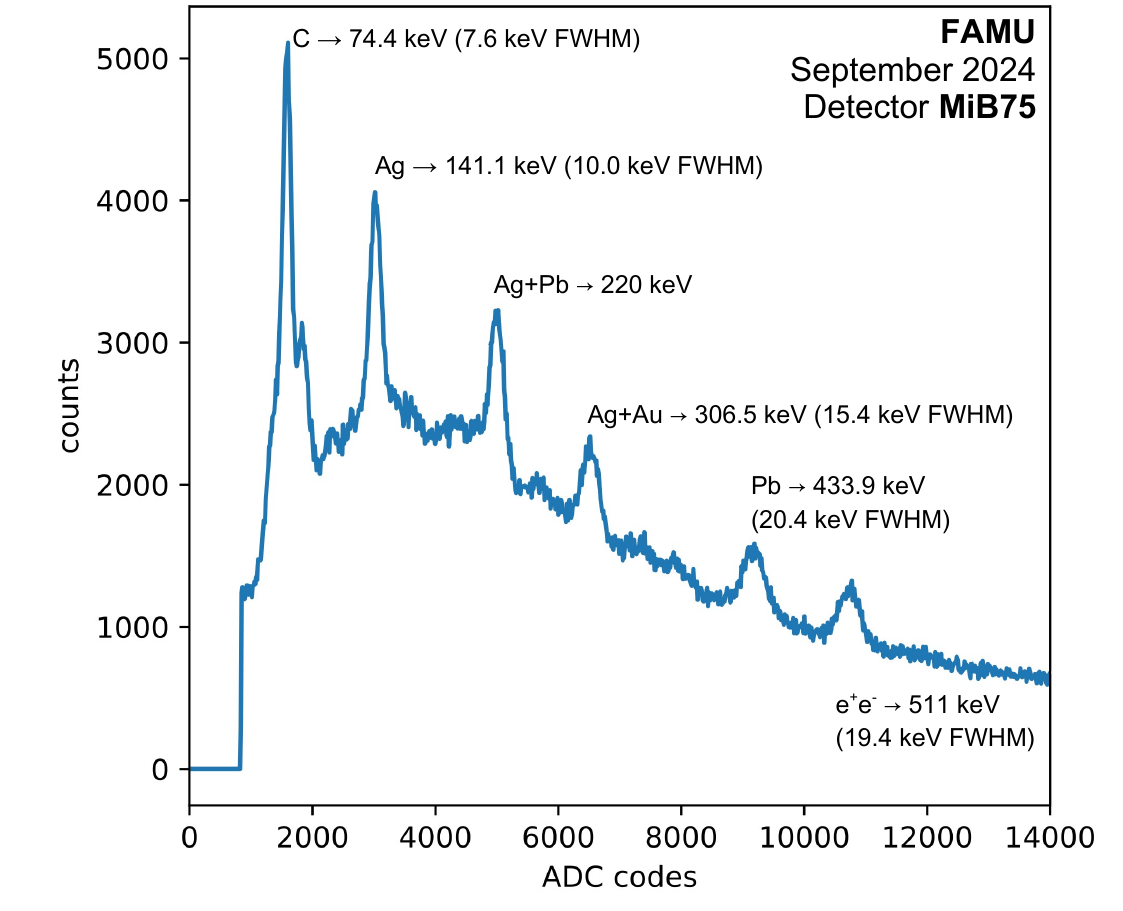}
\caption{Example of prompt X-rays from muon atoms of the material composing the FAMU target. The 511 keV prompt peak of e$^+$e$^-$ annihilation is also highlighted.}
\label{fig:calib_det}       
\end{figure*}

Before each FAMU beam time, all X-ray detectors are calibrated online with radioactive sources of $^{241}$Am (60~keV gamma ray), $^{133}$Ba (multi-$\gamma$ with main peaks at 31~keV, 81~keV, 356~keV) and $^{137}$Cs (662~keV gamma ray). To do so, the muon beam is off and the trigger is given by the detector signals itself. A more refined offline semi-automatic calibration procedure allows then to reconstruct the spectral lines and to evaluate the time-dependence of the gain of each detector through the entire data taking period of the FAMU experiment. The gain variation is monitored by looking at the lines of the prompt muonic captures on the materials composing the target. Fig.~\ref{fig:calib_det} shows an example of energy spectrum in ADC counts for one of the 1" {\em MIB} detectors.  

\begin{figure*}[htpb]
\centering
\includegraphics[width=.49\linewidth]{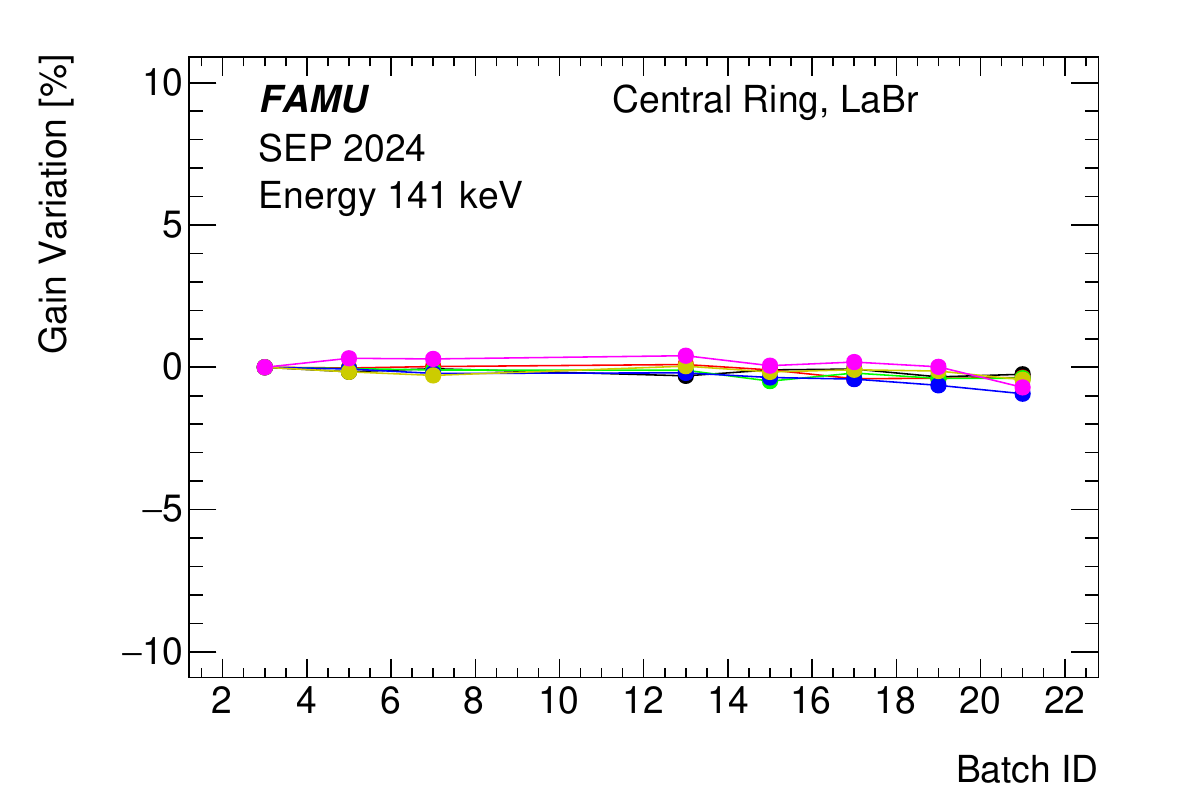}
\includegraphics[width=.49\linewidth]{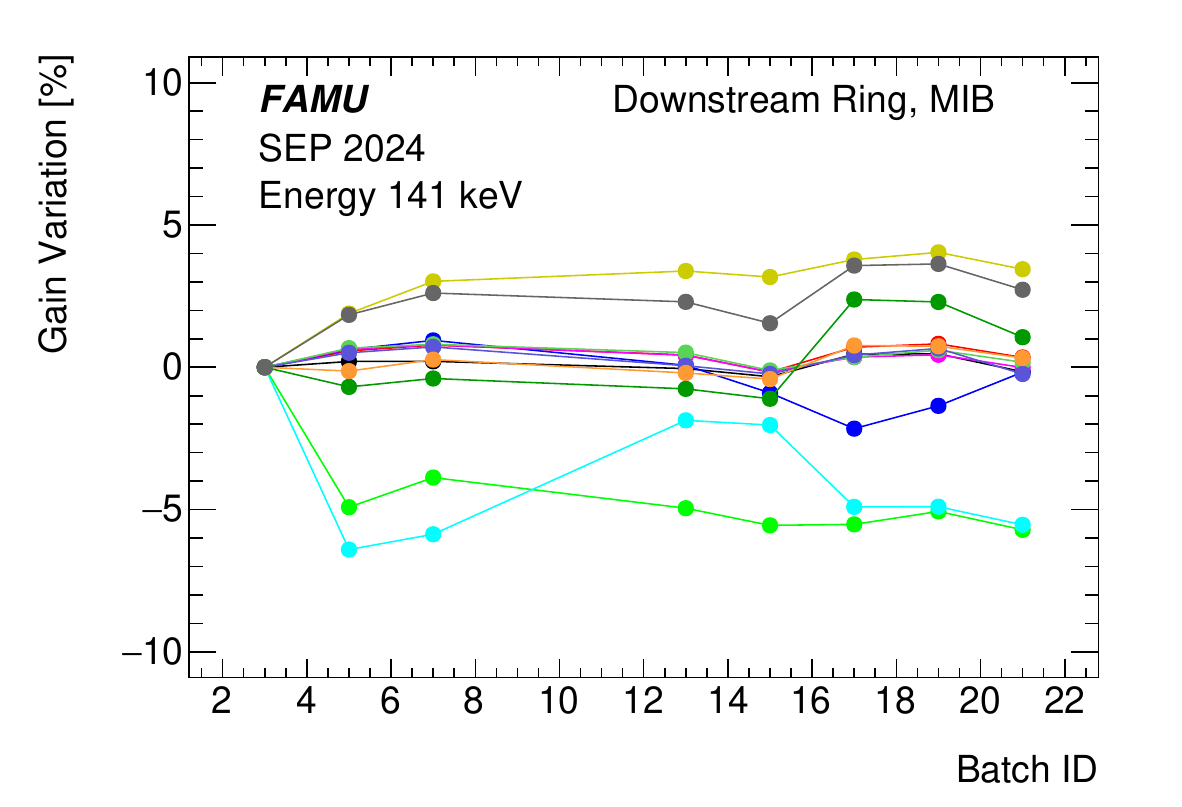}
\includegraphics[width=.49\linewidth]{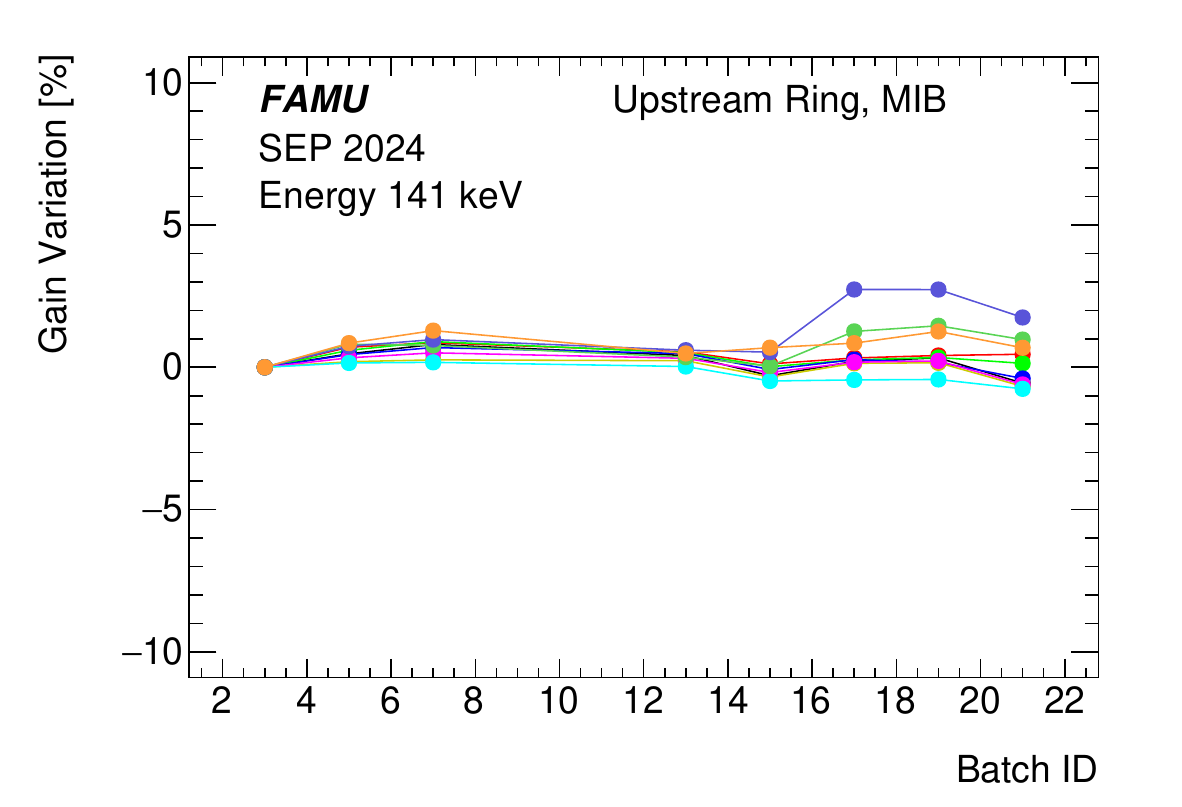}
\includegraphics[width=.49\linewidth]{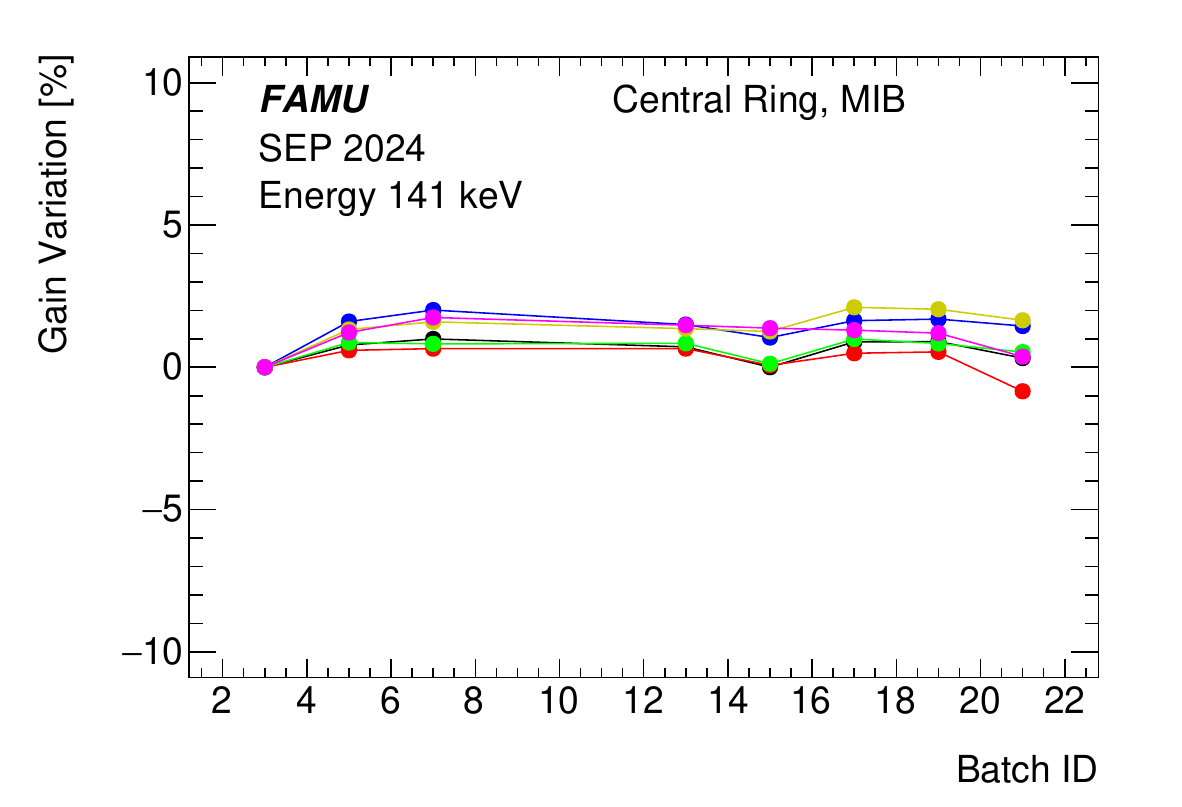}
\caption{Gain variation as a function of the time of LaBr$_3$:Ce detectors with PMT ({\em LaBr}, top-left) and SiPM ({\em MIB}, other panels) readout. The time on the x-axis is expressed in terms of the batch identification number ({\em Batch N}) used to uniquely identify each data sample within the same data taking period. The gain variation is evaluated as percent difference of the muonic silver peak position (E = 141 keV) measured in each sample with respect to first sample (Batch 3). After offline equalization along the different batches, the gain variation for SiPM array readout detectors is compatible with zero.}
\label{fig:gain}       
\end{figure*}

\begin{figure*}[htpb]
\centering
\includegraphics[width=.8\linewidth]{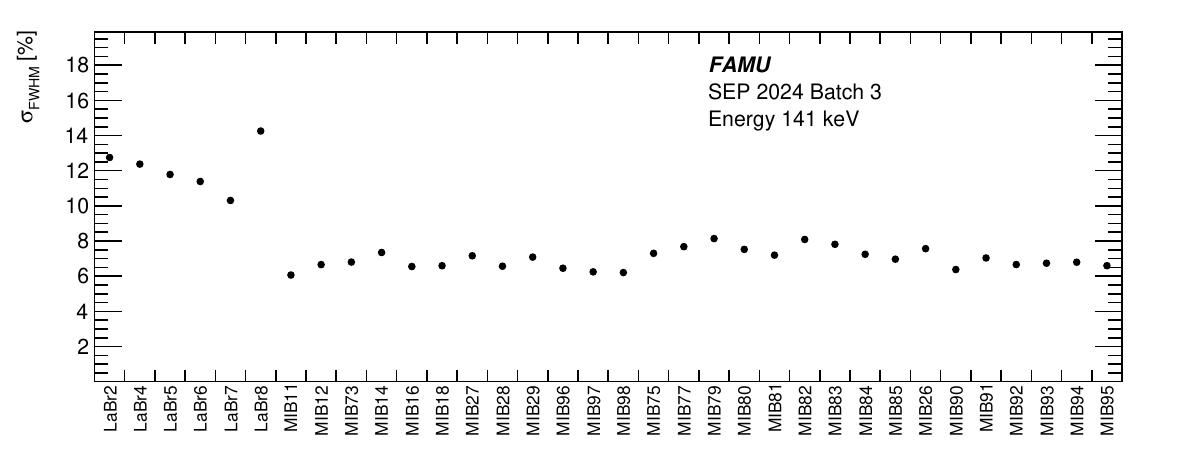}
\caption{FWHM energy resolution at 141 keV as a function of the detector name measured in all 34 detectors used in the September 2024 beam time identified by the Batch 3.}
\label{fig:plotresolution}
\end{figure*}

\begin{figure*}[htpb]
\centering
\includegraphics[width=.49\linewidth]{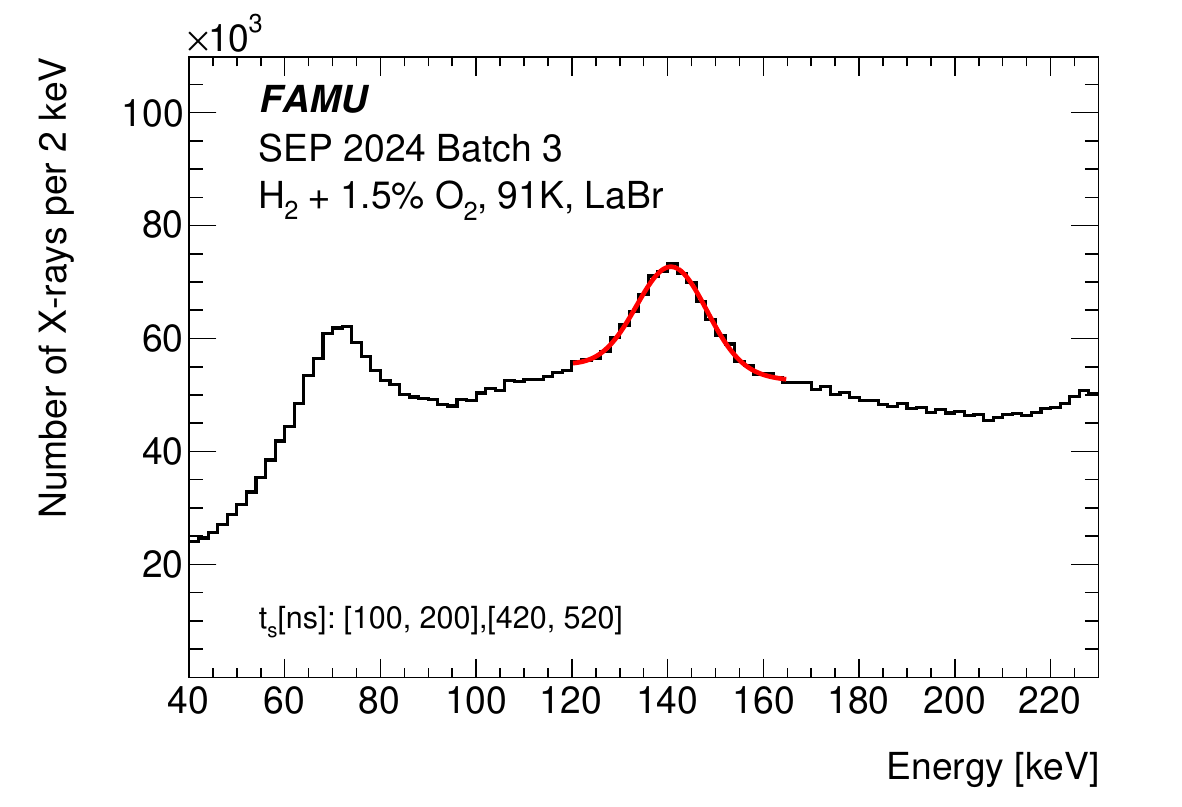}
\includegraphics[width=.49\linewidth]{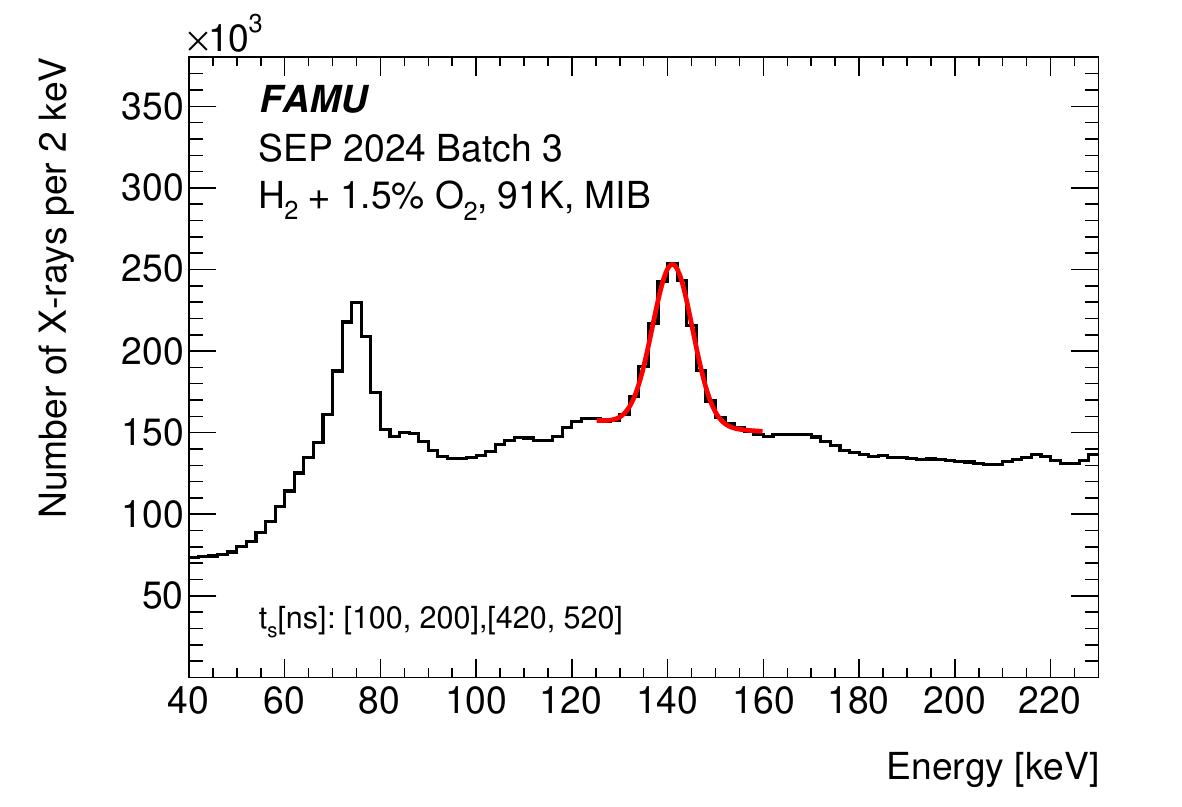}
\caption{Calibrated energy spectra (zoom in the region 40$\div$ 230 keV) obtained by adding up the pulses in all detectors of the same type in the sample with Batch 3. The left plot is obtained using 6 {\em LaBr} detectors, while the right plot results from the combination of 28 {\em MIB} detectors. The two peaks at 74 keV and at 141 keV correspond to the prompt X-ray emission following muon beam absorption in Carbon and Silver, respectively.
As an example, a Gaussian fit to the data is superimposed in the region around 141 keV assuming a 1-degree polynomial background.
The FWHM energy resolution at 141 keV is $12.3 \%$ for the combined {\em LaBr} detectors and $7.0 \%$ for the combined {\em MIB} detectors.}
\label{fig:plotenergy}       
\end{figure*}

As it can be seen in Fig.~\ref{fig:gain}, the gain of almost all detectors is stable at a few percent level. Fig.~\ref{fig:plotresolution} indicates that {\em LaBr} detectors have a larger energy resolution (about 12\%) compared to {\em MIB} detectors (about 7\%), despite their better gain stability in time. 
This is partly due to larger ageing effects on the {\em LaBr} detector crystals since they have been used for a longer time in the FAMU experiment compared to {\em MIB} detector crystals. Fig.~\ref{fig:plotenergy} shows the calibrated energy spectra obtained by adding up the pulses in all detectors of the same type ({\em LaBr} and {\em MIB}). Thanks to the offline calibration procedure, the energy resolution measured in the sample, obtained combining all detectors of the same type, is consistent with the average energy resolution of the single detectors.

\begin{figure*}[htpb]
\centering
\includegraphics[width=.5\linewidth]{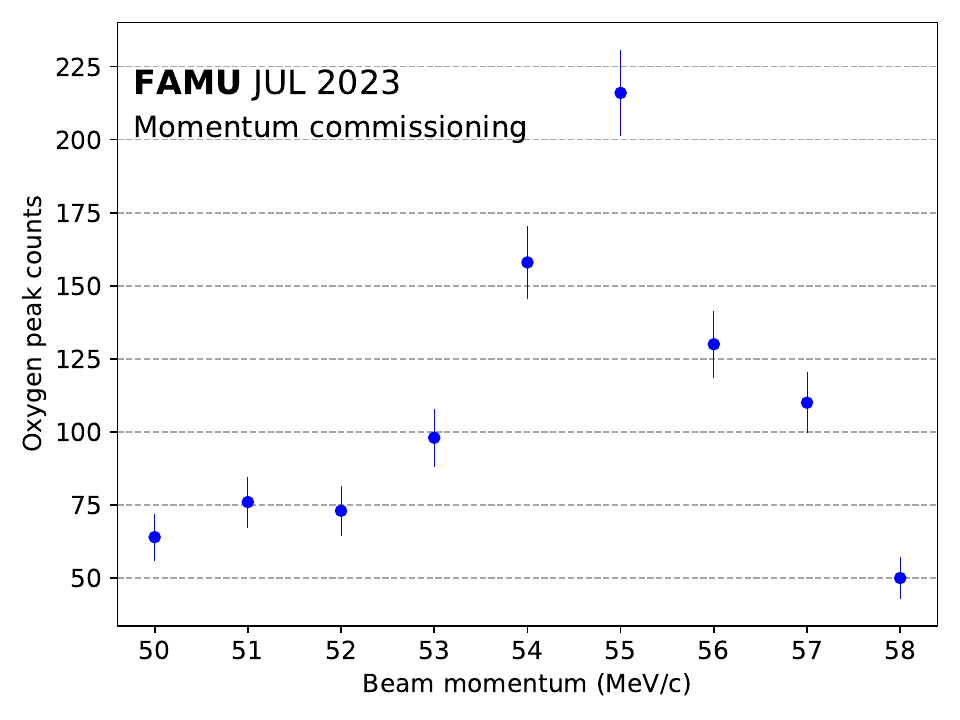}
\caption{Variation of the delayed $\mu$O X-rays as a
function of beam momentum. By selecting p$_\mu$ = 55~MeV/c, the number of muons stopped in the gas is maximised.}
\label{fig:calib_beam}       
\end{figure*}

The optimal muon beam momentum was finally selected in order to maximize the number of muons stopping inside the gas. This FAMU beam commissioning was accomplished by a fine tuning of the magnets of the RIKEN-RAL beam line, followed by the observation of prompt X-rays, coming from the de-excitation of muonic Oxygen atoms, calibrated with the radioactive sources as described above, see Fig.~\ref{fig:calib_beam}.

\subsection{FAMU runs for physics}
\label{subsec:physics}

\begin{figure}[htbp]
        \centering
        \includegraphics[height=0.32\linewidth]{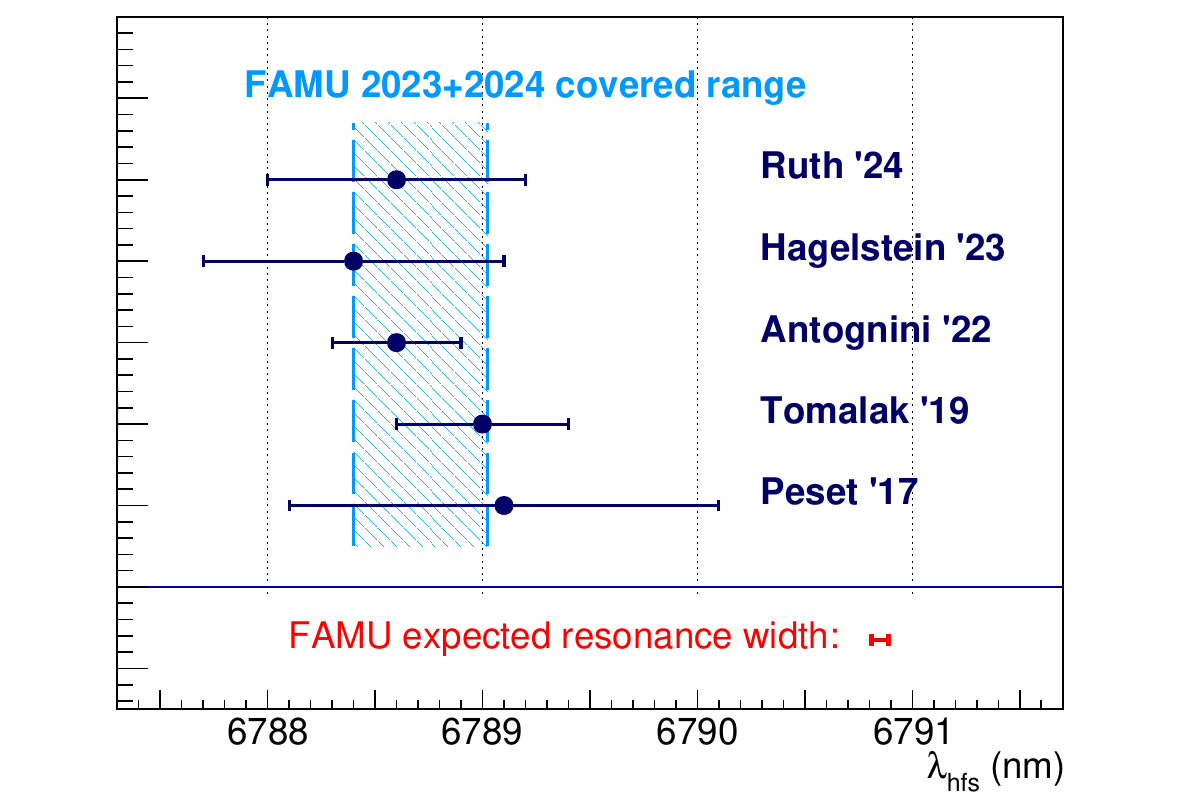}
        \includegraphics[height=0.35\linewidth]{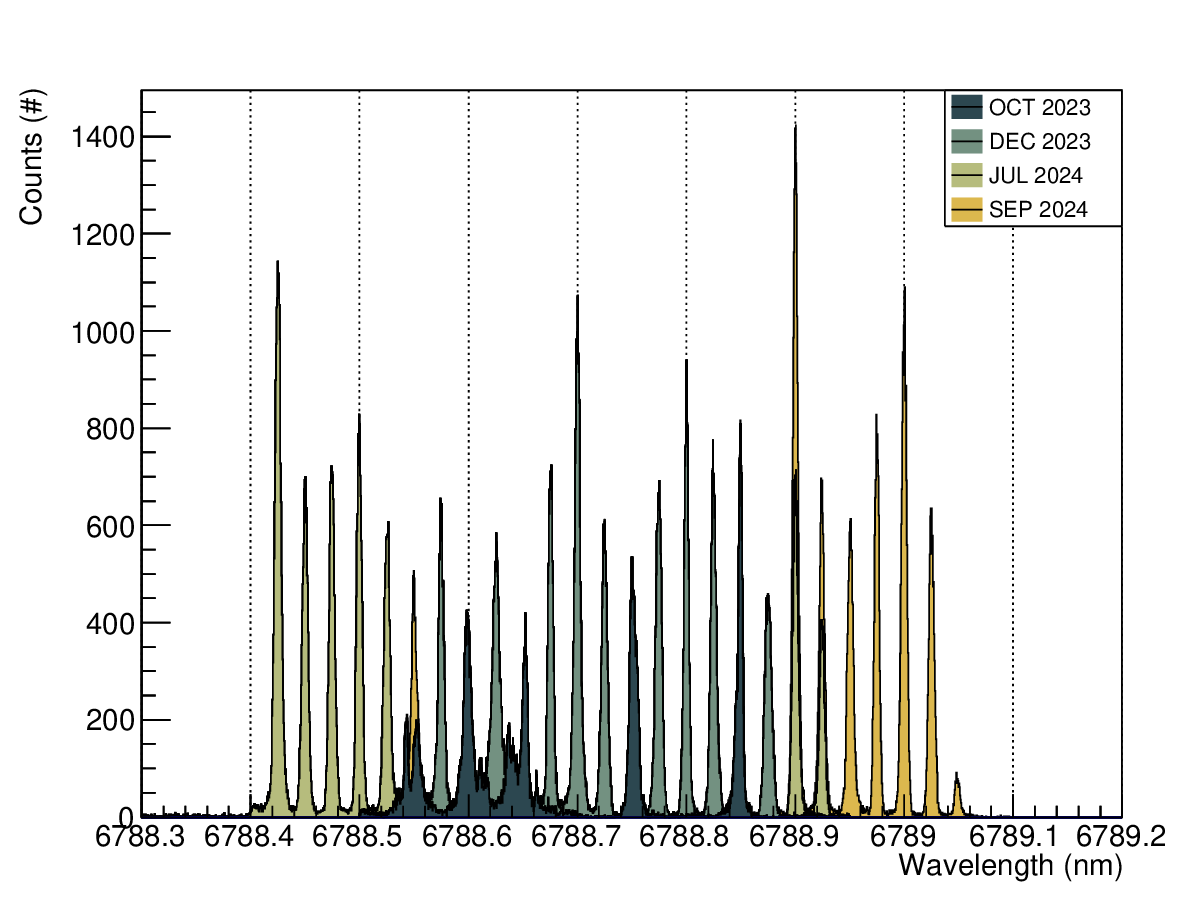}
        \caption{Left: comparison among the latest theoretical predictions for the {\em 1S-hfs} and the spectral range covered in the 2023 and 2024 beam times. Right: histogram of the 29 wavelengths measured in 2023 and 2024 by the FAMU experiment.}
        \label{fig:region}
\end{figure}

The first two beam times for physics took place in October and December 2023 for a total of 17 days. The spectral region covered by these data acquisitions is shown in Fig.~\ref{fig:region} (left) and compared to the latest theoretical predictions for the {\em 1S-hfs} wavelength. In particular, 14 wavelengths have been measured in 2023 with the required statistics (at least 21 – 22 hours of live time) and spanning from 6788.550~nm to 6788.875~nm. The fourth and fifth beam times followed in July and October 2024, respectively, allowing for the further extension of the range of the scanned wavelengths up to 29, from 6788.400~nm to 6789.050~nm. The total statistics of the triggers collected by FAMU in these four periods is reported in Tab.~\ref{tab:statistics}.

\begin{table}[htb]
\label{table-period-summary}
\smallskip
\centering
\begin{tabular}{|c|c|c|c|}
\hline
 Period & Setup &  No. triggers \\
 \hline
 Oct 2023 &  A &  1.60 10$^{7}$  \\
 Dec 2023 &  A &  2.65 10$^{7}$  \\
 Jul 2024 &  B &  2.12 10$^{7}$ \\
 Sept 2024 & B &  2.15 10$^{7}$ \\
\hline
\end{tabular}
\caption{Summary of the data acquisition periods with the total statistics of the collected triggers. Setup A and B are described in Sec.~\ref{subsec:labr}.}
\label{tab:statistics}
\end{table}

The off-line data analysis is a key part for the experiment as it allows to select the information contained in the collected events to extract detector performance and
the data required to construct the plot of the signal as a function of the laser wavelength as already shown in Fig.~\ref{fig:resonance}. Methods developed to get the final resonance plot using the first runs for physics are here briefly described and first preliminary results are shown.

\begin{figure}[htpb] 
            \centering
            \includegraphics[height=0.32\linewidth]{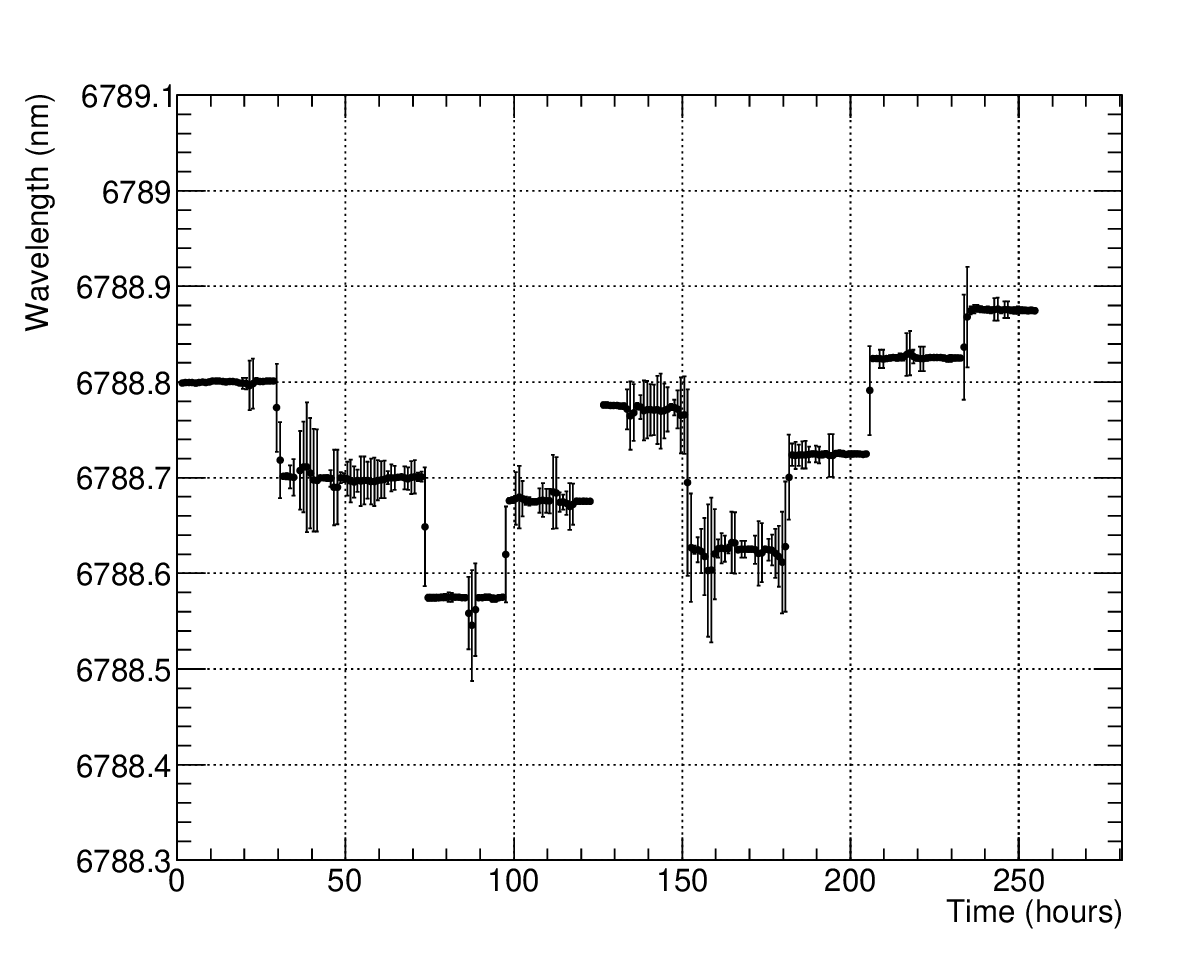}
            \includegraphics[height=0.32\linewidth]{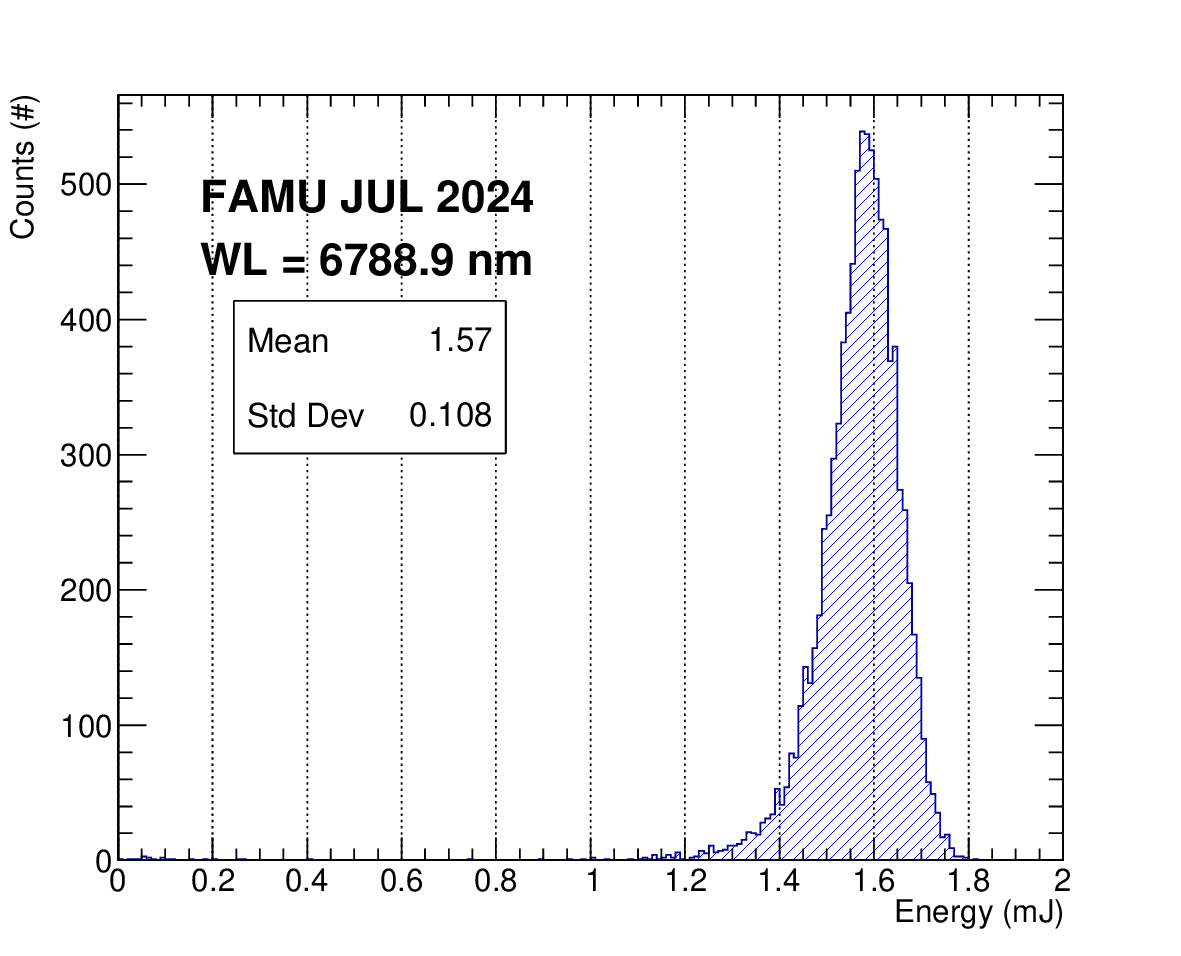}
            \caption{Left: plot of measured wavelengths as a function of time during the December 2023 run, evidencing the excellent stability due to the automatic wavelength keeping software. Right: example of energy distribution during a batch of data acquisition with fixed wavelength.}
            \label{fig:wl}
\end{figure}

The histogram of the laser wavelengths measured during 2023 is shown in Fig.~\ref{fig:region} (right). The histogram is made with the instantaneous value of wavelength saved for each event. The wavelength is varied in steps of 25~pm, as this optimal step allows for scanning the largest portion of the wavelengths without missing the signal, which is expected to be $\sim$ 80~pm wide. It is important to notice that narrower and more regular peaks are visible for December 2023. This is thanks to an automatic wavelength keeper which was added to the laser software to optimise the laser stability. Fig.~\ref{fig:wl} (left) demonstrates the wavelength stability: each point of the plot represents the average of the wavelength over one hour of data taking.

Fig.~\ref{fig:wl} (right) shows a typical distribution of the energy delivered by each laser pulse. It is essential to recognize that the transition probability is proportional to the laser energy, as it is proportional to the number of photons interacting with the $\mu H$ atoms. 
For this reason, the energy distribution must be considered when selecting data and to compare events with different energies. The system is not capable of delivering the same energy for every wavelength and this variation among different batches has to be considered for data normalization.

\begin{figure}[htpb] 
            \centering
        \includegraphics[width=0.5\linewidth]{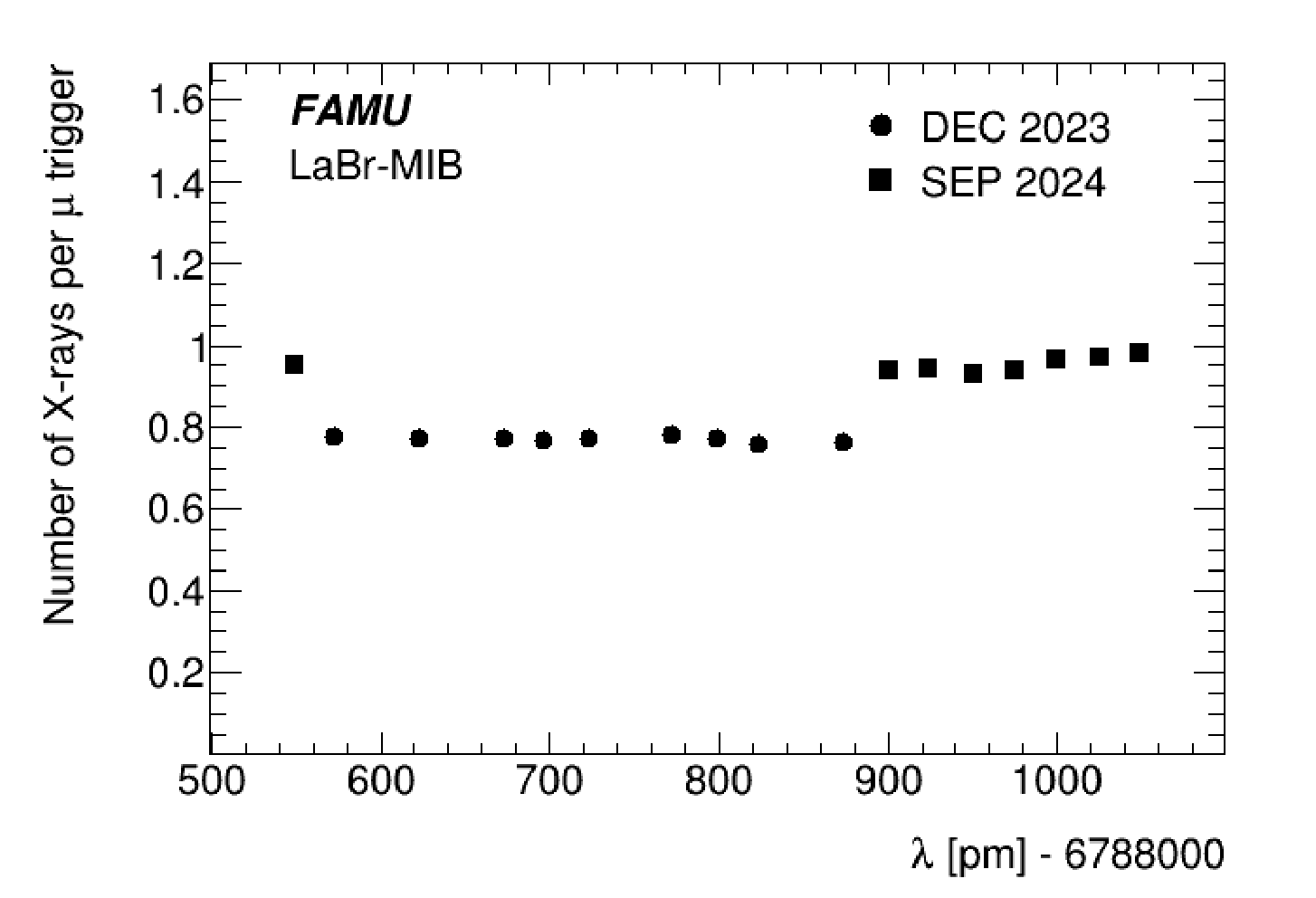}
            \caption{Number of X-rays per muon trigger collected in the Oxygen signal region by all LaBr$_3$:Ce detectors ({\em LaBr}-{\em MIB}), as a function of the wavelength in two different data taking periods.}
            \label{fig:xrays}
        \end{figure}

An example of the performance of the X-ray detectors is shown in Fig.~\ref{fig:xrays}, where the number of X-rays collected in the Oxygen signal region (60$\div$200~keV) by the LaBr$_3$:Ce read by SiPM arrays is shown as a function of the wavelength. The number of X-rays quoted in the plot is normalized on the number of muon triggers. The initial statistics (December 2023), stably around 0.8 X-rays per trigger and per wavelength point, has been then increased up to $\sim$ 1 in the latest run (September 2024) thanks to the increase of the geometrical acceptance of the X-ray detectors due to the substitution of a number of 0.5" detectors with 1" detectors (setup B, see Sec.~\ref{subsec:labr}).

\begin{figure}[htpb]
        \centering
        \includegraphics[width=0.45\textwidth]{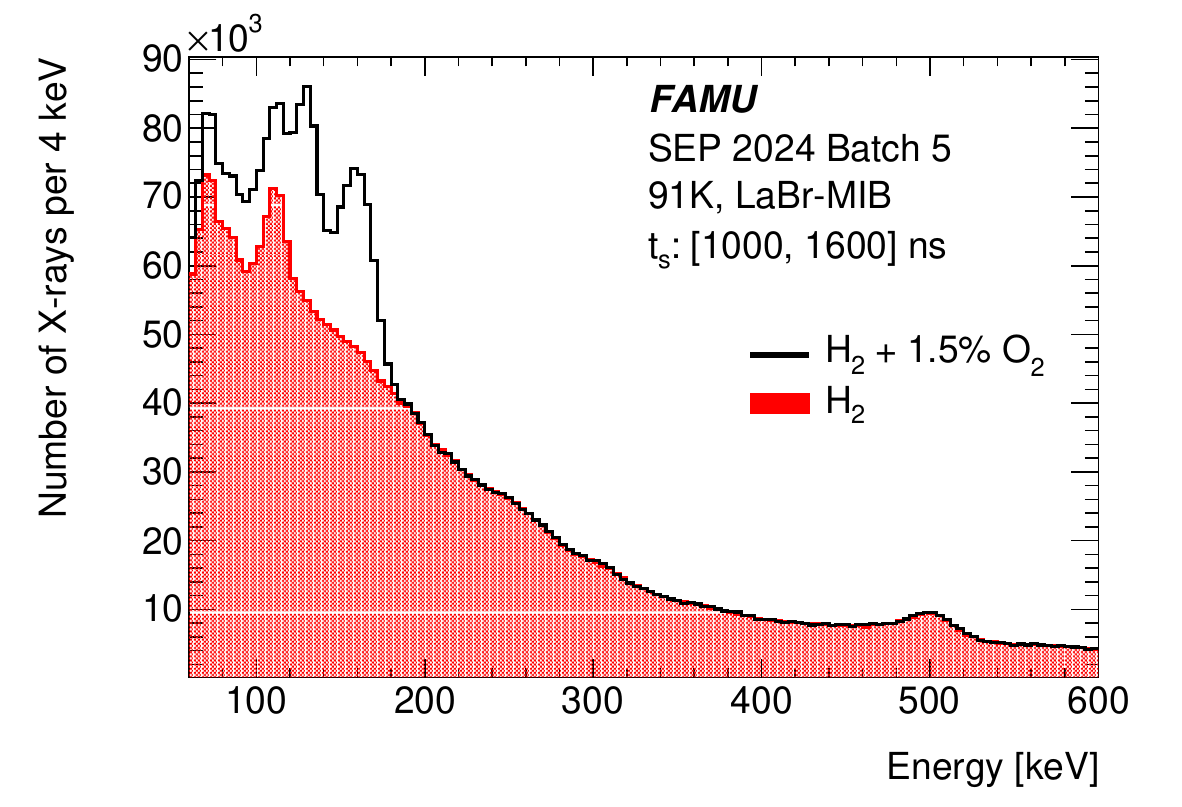}
        \includegraphics[width=0.45\textwidth]{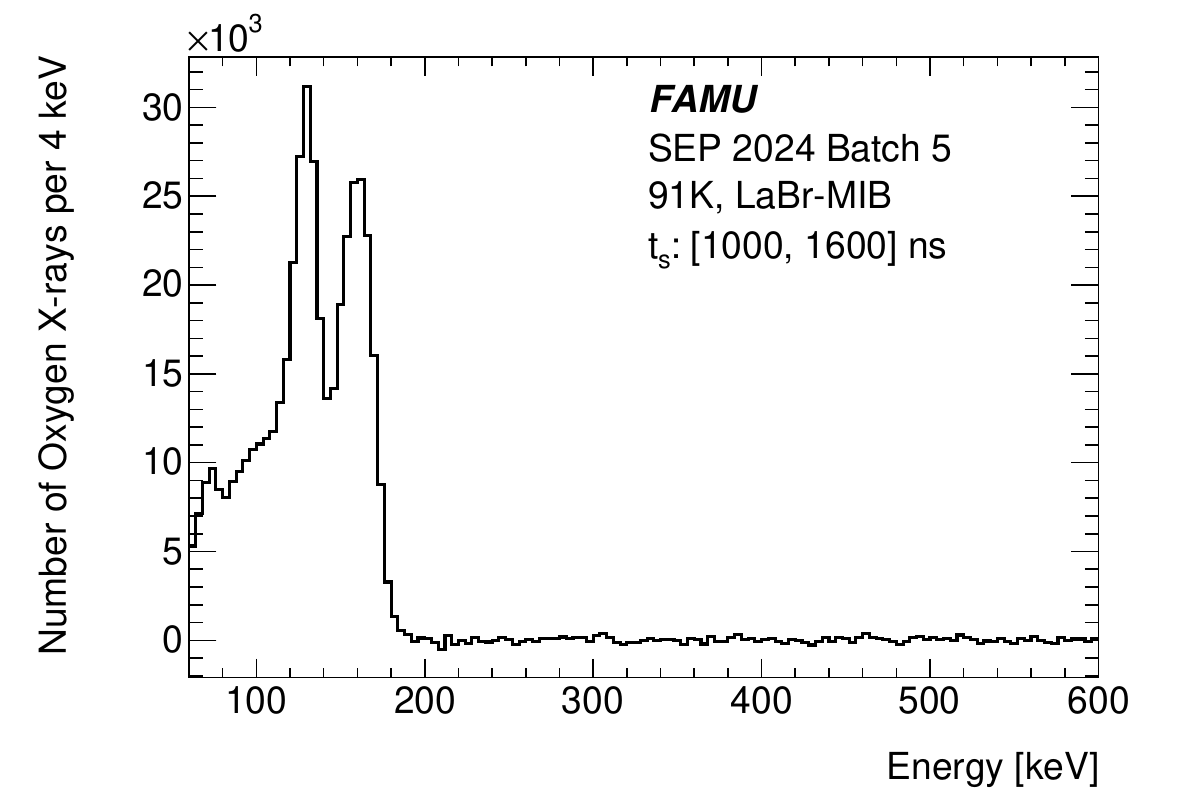}
        \caption{Subtraction of the normalised $H_2$ delayed background, measured in September 2024, from the delayed spectrum in Batch 5, for all LaBr$_3$:Ce detectors ({\em LaBr}-{\em MIB}). The plot on the left shows the two spectra, the plot on the right the net subtraction which is the net oxygen contribution.}
        \label{fig:delayed}
    \end{figure}

Fig.~\ref{fig:delayed} shows with a black line the spectrum measured in 23 hours of data taking measured in September 2024 with all X-ray detectors and the gas mixture ($H_2$+$O_2$). By subtracting this measurement with one carried out with pure hydrogen in the same period (red filled spectrum in the same plot) it was possible to extract the net oxygen contribution (right panel). Events under the K$_\alpha$ and K$_\beta$/K$_\gamma$ oxygen peaks corresponds to X-rays not fully contained in the measuring devices. The origin of the peak at approximately 110 keV, observed in both the signal and background spectra, is under investigation and may plausibly result from muon- or neutron-induced activation of silver or lead within the target.

The following step towards the final {\em 1S-hfs} measurement would then be the evaluation of the total delayed $\mu$O emission rate, separating the collected events in laser and no-laser sub-datasets. In fact, any net signal would have to show up as illustrated in Fig.~\ref{fig:resonance} (left). The actual separation of data in laser and no-laser sub-datasets is currently under investigation, aiming at a blinded analysis protocol. In this way, cognitive biases in the finding of hints of signal in given areas of the spectrum should be avoided. However, before this unblinding can be done, it is still important to set up a testing protocol for systematics in order to make sure that they are not conditioning the collected data. A first result on the {\em 1S-hfs} of the muonic hydrogen from the FAMU Collaboration will then follow.

\section{Conclusions}
\label{sec:conclusions}

The FAMU experiment, designed for a high-precision measurement of the hyperfine splitting in muonic hydrogen, reached its final stage in the summer of 2023 after years of technological and methodological development. For the first time, the final layout it was exposed to the RIKEN-RAL muon beam in its fully operational configuration. All FAMU sub-detectors were thoroughly characterized, and their performance was validated before the start of physics runs, which began in the fall of 2023. A key component of the experiment is the FAMU MIR laser, whose unprecedented characteristics and energy delivery enable the excitation of a large number of muonic hydrogen atoms, significantly increasing the statistics of collected X-rays.

The first physics runs confirmed the flawless operation of the experimental setup, allowing for a precise wavelength scan within the range predicted by the latest theoretical studies for the hyperfine resonance of muonic hydrogen. Following data analysis, the refinement of analytical methodologies, and the evaluation of systematic uncertainties, the FAMU Collaboration will be able to present its first results on the measurement of the hyperfine splitting in muonic hydrogen and, consequently, an assessment of the proton’s Zemach radius. FAMU thus has the potential to open a new window into the proton’s structure, probing it with unprecedented precision.

\section*{Acknowledgements}
Authors would like to thank RIKEN-RAL and ISIS-STFC for the beam time and technical support during experiments (beamtime reference number RB2000022).
We are grateful of the constant support of the INFN CSN3 financing board. We thank Elettra for support on the laser work, and in particular, the SPIE-ICTP Anchor Research Program funded generously by the International Society for Optics and Photonics (SPIE).
Authors are grateful to the Italian Ministry for University and Research (MUR) for supporting the FAMU collaboration through the PRIN 2022 project n. 2022KMNSR9, entitled {\em ``MEtrology and Nonlinear optics for Precision muonic HYdrogen physicS"} (MENPHYS). The research activity presented in this paper has been carried out with the partial support from the Bulgarian Science Fund Grant KP-06-N58/5. The collaborative support of Criotec Impianti srl. in designing, manufacturing, and operating the cryogenic targets that allowed the development of the final layout must be emphasized. We would like to also thank Andrea Abba and Francesco Capogno of Nuclear Instruments for their help in the electronics of the LaBr$_3$:Ce detectors with SiPM readout. The skilful help of the INFN Milano Bicocca and INFN Trieste mechanics workshops for the experimental setup is gratefully acknowledged, as well as the precious contribution by INFN Pavia electronic service, in particular from Marco~C. Prata. We extend our gratitude to Lorenzo Panico of the INFN Naples mechanical workshop for the realization of the two identical multi-pass cavities. 


\begin{thebibliography}{}
%
%

\bibitem{bib:antognini2013}
A. Antognini et al., {\em Proton Structure from the Measurement of 2S-2P Transition Frequencies of Muonic Hydrogen}, Science {\bf 339}, 417–420 (2013).
\bibitem{bib:pohl2013}
R. Pohl et al., {\em Muonic Hydrogen and the Proton Radius Puzzle}, Annu. Rev. Nucl. Part. Sci. {\bf 63}, 175–204 (2013).
\bibitem{bib:bakalov1993}
D. Bakalov et al., {\em Experimental method to measure the hyperfine splitting of muomic hydrogen ($\mu$-p)1S}, Phys. Lett. {\bf A172}, 277–280 (1993).
\bibitem{bib:dupays2003}
A. Dupays et al., {\em Proton Zemach radius from measurements of the hyperfine splitting of hydrogen and muonic hydrogen}, Phys. Rev. {\bf A68}, 052503 (2003).
\bibitem{bib:antognini2022}
A. Antognini et al., {\em The proton structure in and out of muonic hydrogen}, Annual Review of Nuclear and Particle Science {\bf 72} (2022). 
\bibitem{bib:zemach56} 
A. C. Zemach, {\em Proton Structure and the Hyperfine Shift in Hydrogen}, Phys. Rev. {\bf 104} 1771 (1956).
\bibitem{bib:pascalutsa2021}
V. Pascalutsa et al., {\em Theoretical discrepancies in the nucleon spin structure and the hyperfine splitting of muonic hydrogen}, Proceedings of Science CD2021 {\bf 102} (2024).
\bibitem{bib:carlson2011}
C.E. Carlson et al., {\em Proton-structure corrections to hyperfine splitting in muonic hydrogen}, Phys. Rev. {\bf A83}, 042509 (2011).
\bibitem{bib:karshenboim2005}
S. G. Karshenboim, {\em Precision physics of simple atoms: QED tests, nuclear structure and fundamental constants}, Phys. Rep. {\bf 422}, 1 (2005).
\bibitem{bib:adamczak2001}
A. Adamczak et al., {\em On the Use of a H$_2$-$O_2$ Gas Target in Muonic Hydrogen Atom Hyperfine Splitting Experiments}, Hyperfine Interactions {\bf 136}, 1 (2001).
\bibitem{bib:adamczak2012}
A. Adamczak et al., {\em Hyperfine spectroscopy of muonic hydrogen and the PSI Lamb shift experiment}, Nucl. Instr. Meth. {\bf B281}, 72–76 (2012).
\bibitem{bib:pizzolotto2020}
C. Pizzolotto et al. (FAMU Collaboration), {\em The FAMU experiment: muonic hydrogen high precision spectroscopy studies}, Eur. Phys. J. {\bf A56}, 185 (2020).
\bibitem{bib:kanda2018}
S. Kanda et al., {\em Measurement of the proton Zemach radius from the hyperfine splitting in muonic hydrogen atom}, J. Phys. Conf. Ser. {\bf 1138}, 012009 (2018).
\bibitem{bib:amaro2022}
P. Amaro et al., {\em Laser excitation of the 1s-hyperfine transition in muonic hydrogen}, SciPost Phys. {\bf 13}, 020 (2022). 
\bibitem{bib:adamczak2018}
A. Adamczak et al., {\em The FAMU experiment at RIKEN-RAL to study the muon transfer rate from hydrogen to other gases}, JINST {\bf 13}, P12033 (2018).
\bibitem{bib:stoilov2023}
M. Stoilov et al., {\em Experimental determination of the energy dependence of the rate of the muon transfer reaction from muonic hydrogen to oxygen for collision energies up to 0.1 eV}, Phys. Rev. {\bf A107}, 032823 (2023).
\bibitem{bib:antognini2015}
A. Antognini, International Conference on Laser Spectroscopy, ICOLS 2015, Singapore. (2015). arXiv:1512.01765
\bibitem{bib:sato2014}
M. Sato et al., Proceedings of the 20th Particles and Nuclei International Conference, Hamburg, 2014. https://doi.org/10.3204/DESY-PROC-2014-04/67
\bibitem{bib:adamczak2016}
A.~Adamczak et al. (FAMU Collaboration), {\em Steps towards the hyperfine splitting measurement of the muonic hydrogen ground state: pulsed muon beam and detection system characterization} JINST {\bf 11}, P05007 (2016).
\bibitem{bib:vacchi2016}
A.~Vacchi et al. (FAMU Collaboration), RIKEN Accel Prog Rep \textbf{49} (2016).
\bibitem{bib:vacchi2017}
A.~Vacchi et al. (FAMU Collaboration), RIKEN Accel. Prog. Rep. \textbf{50} (2017).
\bibitem{bib:mocchiutti2018}
E.~Mocchiutti et al. (FAMU Collaboration), {\em FAMU: study of the energy dependent transfer rate $\Lambda_{\mu p \rightarrow \mu O}$}, J. Phys. Conf. Ser. {\bf 1138} 012017 (2018).
\bibitem{bib:mocchiutti2019}
E.~Mocchiutti et al. (FAMU Collaboration), {\em First measurement of the temperature dependence of muon transfer rate from muonic hydrogen atoms to oxygen} Phys. Lett. {\bf A384}, 126667 (2020).
\bibitem{bib:pizzolotto2021}
C. Pizzolotto et al. (FAMU Collaboration), {\em Measurement of the muon transfer rate from muonic hydrogen to oxygen in the range 70-336 K}, Phys. Lett. {\bf A403}, 127401 (2021).
\bibitem{bib:hillier2019}
A. D. Hillier et al., {\em Muons at ISIS}, Philosophical Transactions of the Royal Society {\bf A377.2137}, 20180064 (2019).
\bibitem{bib:thomason2019}
J. W. G. Thomason, {\em The ISIS Spallation Neutron and Muon Source - The first thirty-three years}, Nucl. Instr. Meth {\bf A917}, 61 (2019).
\bibitem{bib:carbone2015}
R.~Carbone et al., {\em The fiber-SiPMT beam monitor of the R484 experiment at the RIKEN-RAL muon facility}, JINST {\bf 12}, C03007 (2015). 
\bibitem{bib:bonesini2017}
M.~Bonesini et al., {\em The construction of the Fiber-SiPM beam monitor system of the R484 and R582 experiments at the RIKEN-RAL muon facility}, JINST  {\bf 12}, C03035 (2017).
\bibitem{bib:bonesini2019}
M.~Bonesini et al., {\em The upgraded beam monitor system of the FAMU experiment at RIKEN–RAL}, Nucl. Instr. Meth. {\bf A936}, 592 (2019). 
\bibitem{bib:rossini2024}
R.~Rossini et al. (FAMU Collaboration), {\em The muon beam monitor for the FAMU experiment: design, simulation, test, and operation}, Front. Detect. Sci. Technol. {\bf 2}, 1438902 (2024). 
\bibitem{bib:baruzzo2024}
M.~Baruzzo et al., {\em A mid-IR laser source for muonic hydrogen spectroscopy: The FAMU laser system}, Optics and Laser Technology, {\bf 179}, 111375 (2024).
\bibitem{bib:stoychev2014}
L.I.~Stoychev et al., {\em DFG-based mid-IR laser system for muounic-hydrogen spectroscopy}, Laser Sources and Applications II, Proceeding volume {\bf 9135}, 91350J (2014). 
\bibitem{bib:stoychev2015}
L.I.~Stoychev et al., {\em Increasing the output energy of MID-IR laser system for muounic-hydrogen spectroscopy}, 2015 Fotonica AEIT Ital. Conf. Photonics Technol. Turin Italy May 6-8, 2015
\bibitem{bib:stoychev2019}
L.I.~Stoychev et al., {\em Pulse amplification in a Cr4+:forsterite single longitudinal mode (SLM) multi-pass amplifier}, Laser Phys. {\bf 29}, 065801 (2019).
\bibitem{bib:stoychev2020}
L.I.~Stoychev et al., {\em DFG-based mid-IR tunable source with 0.5 mJ energy and a 30 pm linewidth}, Optics Letters {\bf 45}, 5526 (2020).
\bibitem{bib:baldazzi2017}
G.~Baldazzi et al., {\em The LaBr$_3$(Ce) based detection system for the FAMU experiment}, JINST {\bf 12}, C03067 (2017).
\bibitem{bib:bonesini2022}
M.~Bonesini et al., {\em Large area LaBr$_3$:Ce crystals read by SiPM arrays with improved timing and temperature gain drift control}, Nucl. Instr. Meth. {\bf A1046}, 167677 (2022). 
\bibitem{bib:bonesini2023}
M.~Bonesini et al., {\em One inch LaBr3:Ce detectors, with temperature control and improved time resolution for low energy X-rays spectroscopy}, PoS (EPS-HEP2023), 547 (2023).
\bibitem{bib:bonesini2020}
M.~Bonesini et al., {\em Ce:LaBr3 crystals with SiPM array readout and temperature control for the FAMU experiment at RAL}, JINST {\bf 15} C05065 (2020).
\bibitem{bib:bonesini2021}
M.~Bonesini et al., {\em Detection of low-energy X-rays with 1/2 and 1 inch LaBr3:Ce crystals read by SIPM arrays}, PoS (EPS-HEP2021), 770 (2021).
\bibitem{bib:rossini2024_1}
R.~Rossini et al., {\em Status of the detector setup for the FAMU experiment at RIKEN-RAL for a precision measurement of the Zemach radius of the proton in muonic hydrogen}, JINST {\bf 19} C02034 (2024).
\bibitem{bib:rossini2024_2}
R.~Rossini et al. (FAMU Collaboration), {\em The 2024 LaBr(Ce) detector setup for the FAMU experiment}, Nucl. Instr. Meth. {\bf A1069} 169953 (2024).
\bibitem{bib:bonesini2023_1}
M.~Bonesini et al., {\em Improving the time resolution of large area LaBr3:Ce detectors with SiPM array readout}, Condens. Matter {\bf 8}, 99 (2023).
\bibitem{bib:otte2017}
N.~Otte et al., {\em Characterization of three high efficiency and blue sensitive silicon photomultipliers}, Nucl. Instr. Meth. {\bf A846}, 106 (2017).
\bibitem{bib:bonesini2022_1}
M.~Bonesini et al., {\em Online control of the gain drift with temperature of SiPM arrays used for the readout of LaBr3:Ce
crystals}, JINST {\bf 17}, C10004 (2022).
\bibitem{bib:bonesini2025}
M. Bonesini, {\em The fast X-ray detector system of the FAMU experiment at RAL}, Nucl. Instr. Meth. {\bf A1080}, 170780 (2025).
\bibitem{bib:soldani2019}
M. Soldani et al., {\em High performance DAQ for muon spectroscopy experiments}, Nucl. Instr. Meth. {\bf A936}, 327 (2019).

\end{thebibliography}
%

\end{document}